\documentclass[pre,epsf]{revtex4}

\usepackage{graphicx}
\usepackage{amssymb,amsmath}
\usepackage{amsfonts}
\usepackage{mathrsfs}
\usepackage[retainorgcmds]{IEEEtrantools}

\newcommand{\mn}{\mathbf}
\newcommand{\p}{\bullet}
\newcommand{\h}{\circ}

\newcommand{\matriz}[4]{\left(\begin{array}{cc} #1 & #2 \\ #3 & #4 \end{array}\right)}

\def\<{\langle}
\def\>{\rangle}

\def\beq{\begin{equation}}
\def\eeq{\end{equation}}
\def\barray{\begin{eqnarray}}
\def\earray{\end{eqnarray}}
\def\myi{i}
\def\stroke{\vrule height8pt width0.4pt depth-0.1pt}
\font\upright=cmu10 scaled\magstep1
\def\Cmath{\vcenter{\hbox{\upright\rlap{\rlap{C}\kern
                   3.8pt\stroke}\phantom{C}}}}

\begin{document}

\title{Entanglement of excited states in critical spin chains}

\author{Miguel Ib\'a\~nez Berganza$^1$, Francisco Castilho Alcaraz$^2$, Germ\'an Sierra$^1$   }
\affiliation{
$^1$ Instituto de F\'isica Te\'orica UAM/CSIC, Universidad Aut\'onoma de Madrid, Cantoblanco 28049, Madrid, Spain. \\ $^2$ Instituto de F\'isica de S\~ao Carlos, Universidade de S\~ao Paulo, Caixa Postal 369, S\~ao Carlos, SP, Brazil.}


\begin{abstract}
R\'enyi and von Neumann entropies quantifying the amount of entanglement in ground states of critical spin chains are known to satisfy a universal law which is given by the Conformal Field Theory (CFT) describing their scaling regime. This law can be generalized to excitations described by primary fields in CFT, as was done in reference \cite{Alcaraz2011Entanglement}, of which this work is a completion. An alternative derivation is presented, together with numerical verifications of our results in different models belonging to the $c=1,1/2$ universality classes. Oscillations of the R\'enyi entropy in excited states and descendant fields are also discussed.
\end{abstract}

\maketitle


\section{Introduction}

Entanglement is a fundamental property of quantum states according to which measurements in one of two sub-parts of a quantum system can (immediately) condition the state of the other part, which may be far away in space. This property was at the origin of Einstein's criticisms to quantum mechanics in his EPR article in 1935. Today, entanglement is one of the most studied topics in physics and many research lines are related to this concept. It is the property that makes possible quantum computation, teleportation and quantum information processing \cite{Nielsen2004Quantum}. On the other hand, the irruption of quantum information methods and concepts in the condensed matter physics community has led to one of the most fruitful research areas of the last decade \cite{Amico2008Entanglement}. Measurements of entanglement between spatial or more abstract parts of many-body systems and statistical models in their ground state have been proved to unveil essential information characterizing their phases. After all, quantum correlations, or entanglement, in ground states (gs) of many-body systems are responsible for the onset of coherent phases of matter at zero temperature such as superconductivity and Hall states. Probably the most studied quantity in this context is the amount of entanglement between two spatially separated parts of an extended system. Calling $A$ one of these parts, the reduced density matrix, $\rho_A$, of the ground state under study, is obtained by tracing the degrees of freedom of the complementary of $A$. If $A$ results in a mixed (pure) state, then $A$ and its complementary are said to be in an \textit{entangled} (\textit{product}) state, the amount of entanglement being normally quantified \textit{via} the von Neumann entropy of $\rho_A$:

\beq
S_1(A)=-\mbox{tr }\rho_A \ln \rho_A
\label{S1}
,
\eeq
called \textit{entanglement entropy}. Alternatively, the $n$-th order R\'enyi entropy is used, $n>1$: $S_n(A)=\frac{1}{1-n}\ln\mbox{tr}_A\,{\rho_A}^n$, which contains information on the spectrum of $\rho_A$,  the entanglement entropy being $\lim_{n\to1} S_n$. The interest of these quantities is threefold. Firstly because of the \textit{area law} satisfied by (\ref{S1}) (see references in \cite{Eisert2010Colloquium}), stated in its first version by Bombelli \textit{et al} in 1986 and by Srednicki in 1993 for a free scalar field, and related to the problem of information loss in black holes \cite{Bombelli1986Quantum},\cite{Srednicki1993Entropy}. Roughly speaking, it states that ground states of Hamiltonians with short-range interactions are such that their entanglement entropy is proportional to the area of the hyper-surface separating $A$ from the rest of the system, hence proportional to $\ell^{d-1}$ in $d$ dimensions, if $\ell$ is the linear size of $A$. This important property serves to characterize the very small fraction of the Hilbert space accessibly to ``physical'' ground states. In one spatial dimension (1D), the area law implies that the entropy of massive ground states is bounded and independent of the length $\ell$ of the linear part, or \textit{block}, $A$. Being the entanglement entropy proportional to the minimum amount of information needed to describe the partition $A$, this was used (see references in \cite{Cirac2009Renormalization}) to interpret the efficiency of the Density Matrix Renormalization Group (DMRG) algorithm in massive phases \cite{White1992Density}, and the lack of this efficiency when dealing with gapless phases, a case in which the area law is to be corrected with a logarithmic term in one dimension, leading to an unbounded entropy in infinite systems. In the same spirit, measurements of entanglement are essential in the development of novel \cite{Cirac2009Renormalization}  efficient algorithms for the simulation of quantum many-body systems in one and two dimensions. Entanglement measurements, from a broader perspective, may provide relevant information about the physics of the system under study. Just to give an example that will be cited in the body of the article, the oscillations of the R\'enyi entropy in critical 1D chains have been proved \cite{Calabrese2010Parity} to be related to the Luttinger parameter entering in their continuum description, which, in principle, could be inferred in this way through quantum information measurements in particular finite-size models. \\
\indent
Finally, the interest of entanglement entropy comes also from the fact that the mentioned logarithmic behaviour of the entanglement in 1D has been proved to be a universal property of critical systems, captured, among other properties of entanglement, by their underlying Conformal Field Theory (CFT). Since the work of Polyakov \textit{et. al.} \cite{cft_bpz_nuclphys1984}, conformal symmetry of critical two-dimensional systems has been exploited to infer the universal form of correlators, finite size scaling of energy and momentum, critical exponents and several properties of stochastic evolution of interfaces \cite{Mussardo2009Statistical,Cardy2005SLE}. Moreover, in 1994, Holzhey \textit{et. al.}, and Calabrese \textit{et. al.} in 2004 showed in a CFT context that, in a system of length $N$ and periodic boundary conditions (PBCs), the R\'enyi entropy takes the universal form \cite{Holzhey1994Geometric},\cite{Vidal2003Entanglement},\cite{Calabrese2004Entanglement}:

\beq
S_n^{\rm gs}(\ell) = \frac{c\;(n+1)}{6n}  \ln \left[    \frac{N}{ \pi} \sin \left( \frac{ \pi \ell}{N} \right) \right]+\gamma_n
\label{HLWlaw}
\eeq
where $c$ is the central charge of the CFT describing the system scaling limit  and $\gamma_n$ is a non-universal constant. This statement was generalized to account not only for the ground state entropy but also for low-energy excitations in reference \cite{Alcaraz2011Entanglement}, a generalization which supposes a further prediction of CFT in the field of quantum critical phenomena. By conformal invariance, the finite-size spectrum  at criticality has a  universal structure determined by the Virasoro algebra satisfied by the generators of the conformal transformations. Each state in the low-energy spectrum exhibits an energy and momentum finite-size scaling determined by the \textit{conformal weights}, or, roughly speaking, the eigenvalues of the Virasoro operators of the given state. In the same spirit, in reference \cite{Alcaraz2011Entanglement} it was proved that the R\'enyi entropy of excitations defined by primary fields are universally related to conformal properties of the operator defining the targeted excitation. In particular, consider the quantity:

\barray
F^{(n)}_\Upsilon \equiv \exp\left[(1-n)(S_n^\Upsilon(x)-S_n^{\rm gs}(x))\right]
\label{Fndefinition}
,
\earray
defined in reference \cite{Alcaraz2011Entanglement}. It quantifies the excess of entanglement of a state $|\Upsilon\>$ related to the primary field $\Upsilon$, with respect to the gs. This quantity \cite{Alcaraz2011Entanglement} related to the $2n$-point correlator of the operator $\Upsilon$ and its conjugate $\Upsilon^\dag$ in the cylinder:

\barray
F^{(n)}_\Upsilon(x)  = n^{-  2 n ( h + \bar{h})}\,
 \frac{ \langle  \prod_{j=0}^{n-1}  \Upsilon(\frac{2 \pi j}{n}) \, \Upsilon^\dagger( \frac{2 \pi (j + x)}{n})
 \rangle_{\rm cy}}{  
  \langle \Upsilon(0) \, \Upsilon^\dagger( 2 \pi x) \rangle_{\rm cy}^n}    \label{Fn} 
.
\earray
\indent
This is the main result we will alternatively derive in the following section, and that will be numerically checked in subsequent sections for different models.\\
\indent
Entanglement of excited states has  been also considered  in  \cite{Alcaraz2008Finitesize}, where universal scaling of the  negativity of excitations in the $XXZ$ critical chain  was shown. In \cite{Masanes2009Area}  it was shown that a violation of the area law should be expected for the low lying excited states of critical  chains, and in \cite{Alba2009Entanglement}  it was considered the entanglement of very large energy excitations in the $XY$ and $XXZ$ models. In \cite{Calabrese2011Entanglement},\cite{Calabrese2011Entanglementbis}, the law (\ref{Fn}) was applied to study systems with continuous degrees of freedom, where it turned to be an accurate description of the finite-size system entanglement.\\
\indent 
The aim of this work is to complete the results of reference  \cite{Alcaraz2011Entanglement}. In section \ref{XXoverview} an overview of entanglement of excitations in the $XX$ model is given. Section \ref{CFT} is a derivation of our results in the framework of the Calabrese and Cardy computation \cite{Calabrese2004Entanglement} (in \cite{Alcaraz2011Entanglement}, the approach of \cite{Holzhey1994Geometric} was used instead). Section \ref{bosons} presents some of our results for the bosonic theory and some numerical illustrations for three models in this universality class: the $XX$ (free fermions), $XXZ$ and excluded-volume-$XX$ spin chains. The section concludes with a study of the R\'eny entropy oscillations in this theory. Section \ref{Ising} is a study of the free Majorana CFT with numerical tests in the quantum critical Ising chain. A discussion about the entropy of descendant fields is given in section \ref{descendant}.

\section{A first sight on entanglement of excitations in the $XX$ model}\label{XXoverview}

As a first illustration of the behaviour exhibited by block entanglement in excited states of spin chains, let us consider the $XX$ model for $N$ interacting spins occupying the positions of an $N$-site lattice and whose interaction is defined by the Hamiltonian:

\beq
H_{XX}=-\frac{1}{2}\sum_{j=1}^N\left( \sigma_j^x\sigma_{j+1}^x + \sigma_j^y\sigma_{j+1}^y \right),
\label{XX}
\eeq 
where $\sigma_j$ are the spin Pauli matrices acting on the $j$-th spin. This well-known model \cite{Lieb1961Two} is a paradigm of spin chain describing quantum magnetism, it also emerges as the strong on-site repulsion limit of the boson Hubbard model. As discussed in appendix \ref{XYfermionization} and in reference \cite{Lieb1961Two}, the $XX$ model is integrable and it can be mapped, through a Jordan-Wigner transformation, into a problem of lattice free fermions. According to such a mapping, the spectrum of the Hamiltonian (\ref{XX}) coincides with the spectrum of a free-fermionic Hamiltonian:

\beq
H=\sum_{j\in  \Omega} E_j\, d_j^\dag d_j + \rm{constant}
\label{Hfermions}
\eeq 
where $\{d_j,d^\dag_{j'}\}=\delta_{j,j'}$, are the creation and destruction operators of fermions with momentum $k_j=2\pi j/N$, $E_j=-\cos k_j$ is the free-fermionic dispersion relation, and $\Omega$ is a set of $N$ integers or half-integers such that the resulting momenta $k_j$ are the ones determined by the periodicity or anti-periodicity of the boundary conditions in the fermionic problem. In other words: each one of the $2^N$ eigenstates of (\ref{XX}) is associated with one of the $2^N$ fermionic eigenstates of (\ref{Hfermions}):

\beq
\prod_{j=1}^{n_F \le N} d_{m_j}^\dag |0\>
\label{fermionicstate}
\eeq 
(being $|0\>$ the fermionic vacuum annihilated by the $d$'s and $\{m_j\}_{j=1}^{n_F}\equiv \mn K \subset \Omega$, one of the $2^N$ subsets of $\Omega$), and the association is such that the eigenvector in correspondence with the momentum set $\mn K$ has energy eigenvalue: $E=-\sum_{j \in \mn K}\,\cos k_j$. As explained in the appendix, the Jordan-Wigner transformation is such that a fermion Hamiltonian with  anti-periodic boundary conditions (APBCs) is related to the Hamiltonian (\ref{XX}) with periodic boundary conditions (PBCs) for even $n_F$, and with APBCs for odd $n_F$. We will impose APBCs to the fermions. The corresponding $\Omega$ is:

\beq
\Omega = \left\{ \pm \frac{1}{2},\pm \frac{3}{2},\ldots,\pm \frac{N-1}{2} \right\}
.
\eeq

\begin{figure}[h]
\begin{center}
\includegraphics[height=7cm]{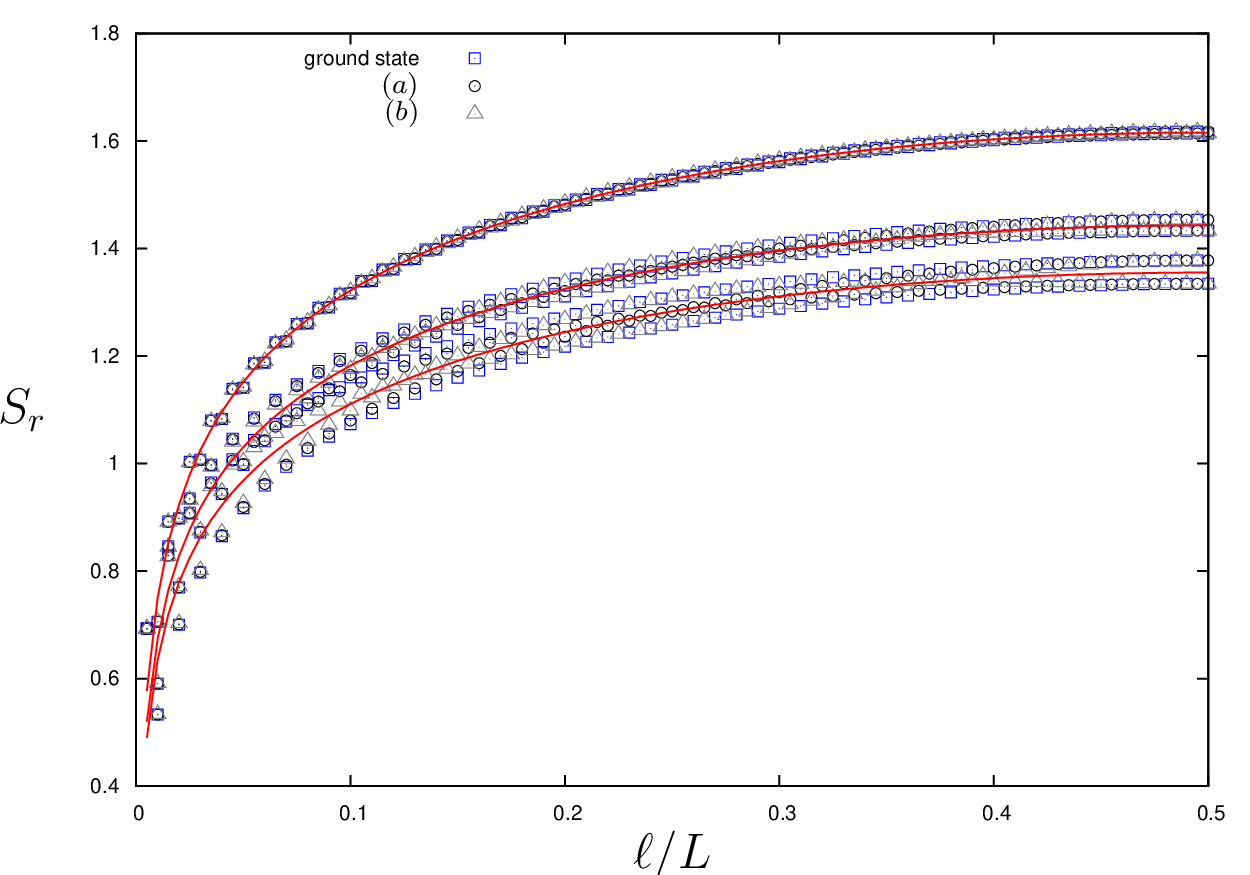}
\caption{\footnotesize R\'enyi entropy $S_r$, $r=2,3,4$ for three states in the $XX$ model with $N=200$ sites. The states $(a)$ and $(b)$ are described in the text and in table \ref{mytable}. The continuous curves are the universal function (\ref{HLWlaw}).}
\label{fermions-1stexcs}
\end{center}
\end{figure}

\indent
In this framework, one has that the ground state of (\ref{XX}) corresponds to the \textit{Fermi state}  of (\ref{Hfermions}) with $n_F=N/2$ fermions:

\barray
|n_F\>= \prod_{0<j\, \le\, (n_F-1)/2} d^\dag_{j}d^\dag_{-j} |0\> 
\label{Fermistate}
.
\earray
%
\indent
From now on we will denote $|m\>$ the Fermi state of $m$ fermions, as in equation (\ref{Fermistate}) (even $m$ is supposed). Let us now consider some low-energy excitations in this model. Due to the conformal invariance, the low-energy excitations present an excess of energy with respect to the ground state which is: $2\pi\Delta/N$, being $\Delta$ the conformal dimension of the excitation. The excess of energy vanishes in the large-$N$ limit: the model is critical. As a critical model, its ground state $|N/2\>$ satisfies the law (\ref{HLWlaw}), one can ask whether the entropy of low-energy excitations coincides with the gs entropy in the thermodynamic limit, as happens for the energy. It turns out that there is a class of low-energy excitations for which this is indeed true. This type of excitations will be called \textit{compact} since, as the ground state, they do not exhibit holes in momentum space or, in other words, for them the set $\mn K$ is composed of consecutive momentum quanta differing by $2\pi/N$. Otherwise the excitations will be called \textit{non-compact}, and they are such that their entropy is larger than that of the ground state. In what follows we illustrate with examples how excitations in the compact class presents an entropy equal to the gs entropy, up to the oscillations described in \cite{Calabrese2010Parity}, with the same non-universal constant $\gamma_n$. Afterwards, some non-compact states will be analyzed. These results will be justified with CFT arguments in section \ref{bosons}.\\
\indent As a first example of compact states we consider the excitation obtained by removing the highest momentum fermion in the $XX$ model, or $d_{n_F/2-1/2}|n_F>$, with $n_F=N/2$. In figure \ref{fermions-1stexcs} we present the $n=2,3,4$-R\'enyi entropy of such a state (which is labelled as ($a$)), together with the ground state entropy. The different behaviour of the oscillations (which are present only for $n>1$, and absent in the $N\to\infty$ limit) is the only difference between ground and excited state entropies. As a further illustration, we present the state obtained adding a fermion below the left Fermi point and another one above the right Fermi point (i.e., the state $d^\dag_{n_F/2+1/2}\, d^\dag_{-n_F/2-1/2} |n_F\>$). This state is labelled as $(b)$ in figure \ref{fermions-1stexcs}. A final example of compact states is provided by the \textit{Umklapp} excitation (c.f. table \ref{mytable}), consisting in moving the fermion with most negative momentum to the right of the positive Fermi point, i.e. to $d_{-(n_F-1)/2}\,d^\dag_{(n_F+1)/2}|n_F\>$. This can be proved\cite{Alcaraz2011Entanglement} to have exactly the same entropy $S_n$ as the ground state for all values of $n$ (with exactly the same excitations), since such a shift in momentum space $k\to k +2\pi/N$ amounts to a phase shift of the wave-function in position space, leaving the reduced density matrix unchanged.\\ \indent For the sake of clarity let us introduce the following notation: $(h_1\,h_2\,\cdots\,:p_1\,p_2\,\cdots)$ refers to a chiral excitation with holes in the $h_j$'s allowed momentum values below the right Fermi point and particles in the $p_j$'s momentum values above it, i.e., to the state:

\barray
\prod_{j,k} d^\dag_{\frac{n_F-1}{2}+p_j}d_{\frac{n_F+1}{2}-h_k} |n_F\> \nonumber
,
\earray
in such a way that the ground state is denoted as $(\,:\,)$ and the ($a$) excitation is $(1:\,)$.\\

\begin{table}
\begin{tabular}{l|c|c|c|c|l}
 name of excitation & field & $(h,\bar h)$ &  state ($n_F=N/2$) &  $(h:p)$  &  $N=8$ example\\
\hline
\small{ground state} & $\mathbf{1}$ & (0,0) & $|n_F\>$ & $(\,:\,)$  & $\h\h\p\p\p\p\h\h$\\
$(a)$ & $e^{-\myi\phi}$  & (1/2,0)  & $d_{(n_F-1)/2}|n_F\>$ &  $(1:\,)$    &           $\h\h\p\p\p\h\h\h$ \\
$(b)$ & $e^{\myi\phi+\myi\bar\phi}$ & $(1/2,1/2)$ & $d^\dag_{n_F/2+1/2}\, d^\dag_{-n_F/2-1/2} |n_F\>$ & & $\h\p\p\p\p\p\p\h$ \\
\small{Umklapp} & $e^{\myi\phi-\myi\bar\phi}$ & $(1/2,1/2)$ &$d_{-(n_F-1)/2}\,d^\dag_{(n_F+1)/2}|n_F\>$ &  & $\h\h\h\p\p\p\p\h$ \\
\hline
\small{particle-hole} & $\myi \partial\phi$ & $(1,0)$ &$d_{(n_F-1)/2}\,d^\dag_{(n_F+1)/2}|n_F\>$ & $(1:1)$  & $\h\h\p\p\p\h\p\h$ \\
\small{R-L particle-hole} & $\bar\partial\bar\phi\partial\phi$ & $(1,1)$ & $\substack{d_{(n_F-1)/2}\,d^\dag_{(n_F+1)/2}\\d_{-(n_F-1)/2}\,d^\dag_{-(n_F+1)/2}|n_F\>}$ &  & $\h\p\h\p\p\h\p\h$ \\
 &  - & - &$d_{(n_F-1)/2}\,d^\dag_{(n_F+3)/2}|n_F\>$ & $(1:2)$  & $\h\h\p\p\p\h\h\p$ \\
\hline
\end{tabular}
\caption{\footnotesize{A summary of the mentioned excitations. The horizontal line separates the compact states from the non-compact ones. The notation $(h,p)$ applies only for chiral excitations, and the corresponding conformal fields are shown for primary states only. }}
\label{mytable}
\end{table}

\begin{figure}[h]
\begin{center}
\includegraphics[height=7cm]{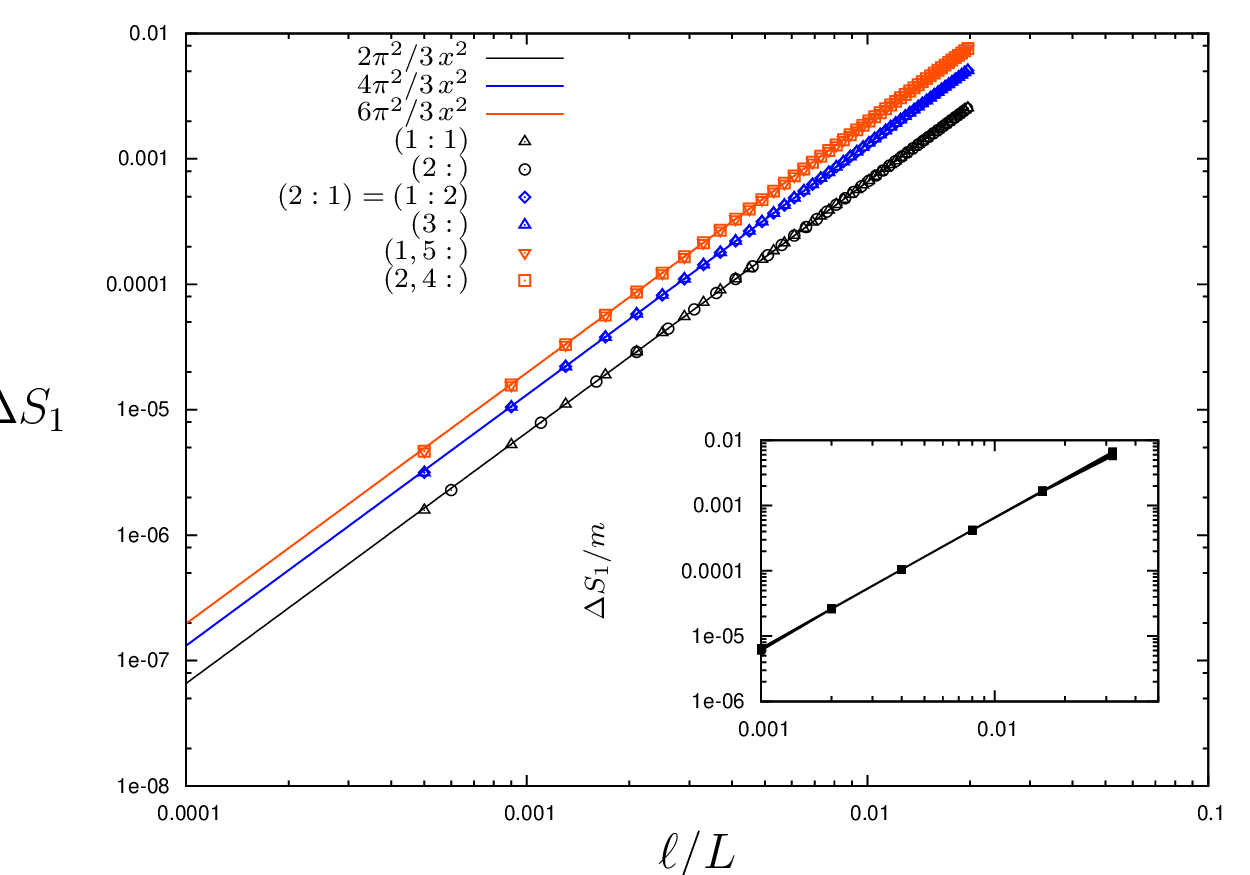}
\caption{\footnotesize Low-$\ell/N$ behaviour of the excess of von Neumann entropy of different states with $\bar h=0$ in the $XX$ model. Continuous lines are equation (\ref{lowxS}). States with the same value of $h$ have a common color. System sizes are $N=10^4$. Inset: several states with $h=m$ ($m$ from 1 to 14) of a $N=8000$ system are shown to satisfy (\ref{lowxS}). }
\label{fermions-lowx}
\end{center}
\end{figure}

\begin{figure}[h]
\begin{center}
\includegraphics[height=7cm]{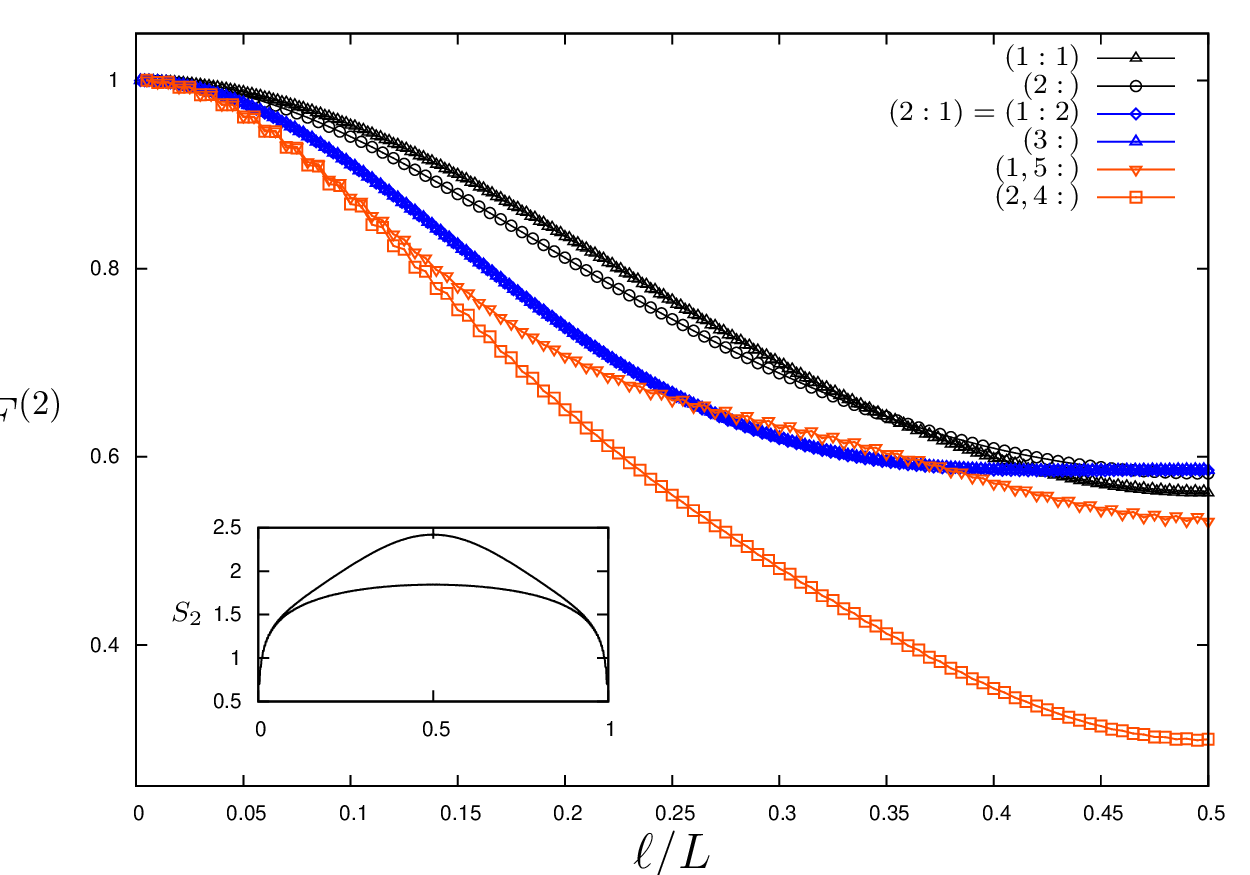}
\caption{\footnotesize The function $F^{(2)}$ for the same excitations of figure \ref{fermions-lowx} in systems with $N=500$ sites. The inset shows $S_2$ for the ground and (1:1) states. }
\label{fermions-extended}
\end{center}
\end{figure}

We will now focus in non-compact excitations, such that holes are created in momentum space. These states do not obey the law (\ref{HLWlaw}). The lowest excitation of the $XX$ model with non-zero momentum has excess of energy and momentum equal to $2\pi/N$, and in the fermionic language it corresponds to the destruction of the fermion below the Fermi point and the creation of a fermion immediately above it. In other words, it is the $(1:1)$ state. We will call this excitation a \textit{particle-hole excitation}. This state is found to exhibit excess of entanglement $\Delta S(\ell)=S(\ell)-S^{\rm gs}(\ell)$ which is larger than zero. Figure \ref{fermions-lowx} represents the $\ell/N << 1$ regime of the excess of entanglement entropy for such a state, together with other particle-hole like excitations. Since the works \cite{Holzhey1994Geometric,Vidal2003Entanglement,Calabrese2004Entanglement}, one knows that, for small $\ell/N$, the ground state of a critical system satisfies $S_1(\ell)\sim \frac{c}{3}\,\ln \ell$. For excited states in the $XX$ model, one observes a correction of the type \cite{Alcaraz2011Entanglement}:

\barray
\Delta S_1(\ell)=\frac{2\pi^2}{3}\,(h+\bar h) \,\left(\frac{\ell}{N}\right)^2+O\left[(\ell/N)^4\right]
\label{lowxS}
\earray
where $\Delta S_1$ is the excess of entropy with respect to the ground state and $h$, $\bar h$ are the conformal weights of the  operator corresponding to the excitation (equation (\ref{lowxS}) will be justified in section \ref{bosons}). As a further example consider the state $ d_{(n_F-1)/2}\, d^\dag_{(n_F-1)/2+m} |n_F\>$, such that the fermion nearest to the Fermi point has been displaced $m$ momentum quanta to the right. On inset of figure \ref{fermions-lowx}, one observes for this state how curves for different $m$ collapse when the quantity $\Delta S_1/m$ is plotted, confirming also in this case the law (\ref{lowxS}) with $h=m$, $\bar h =0$. \\
\indent
It is worth to stress that although some of the states studied in figure \ref{fermions-lowx} present the same low-$x$ dependence of their entanglement and R\'enyi entropies, their entropy do not coincide for larger values of $x$ in general. This is illustrated in figure \ref{fermions-extended}, in which we show the function $F^{(n)}$ defined in (\ref{Fndefinition}) for the excitations studied in figure \ref{fermions-lowx}. \\ \indent In section \ref{bosons} we will show how some of the results presented in this section can be justified with CFT arguments.

\section{CFT approach to the entropy of primary fields} \label{CFT}

\indent 
In reference \cite{Alcaraz2011Entanglement} we obtained a general formula for the R\'enyi entropy of excited states in CFT following the approach of Holzhey,  Larsen and Wilczek  (HLW) who computed  the entanglement entropy of the ground state of CFT  systems  with periodic boundary conditions. 
The HLW  approach was later generalized by Cardy and Calabrese (CC) to deal with more general situations, as finite temperature, open boundary conditions and several disjoint intervals. It is thus of great interest to re-derive the result of reference \cite{Alcaraz2011Entanglement} using the CC approach. This is the aim of this section.\\ \indent Let us start by briefly recalling the main steps of the CC approach. 
The general set up of the problem is a lattice 1+1 quantum field theory  with local commuting  observables $\{ \hat{\phi}(x) \}$  with  eigenvalues  $\{ \phi(x) \}$   and Hamiltonian $\hat{H}$. The  thermal state $\rho$ at inverse temperature $\beta$ has matrix elements

\beq
\langle  \{ \phi''(x'') \} | \rho |  \{ \phi'(x') \} \rangle  = \frac{1}{Z(\beta)} \langle  \{ \phi''(x'') \} | e^{ - \beta \hat{H} }  |  \{ \phi'(x') \} \rangle 
\label{cc5}
\eeq
where

\beq
Z(\beta) = {\rm Tr} \, e^{ - \beta H}
\label{cc6}
\eeq
is the partition function. Equation (\ref{cc5}) can be expressed in terms of an euclidean path integral as

\beq
\langle  \{ \phi''(x'') \} | \rho |  \{ \phi'(x') \} \rangle   = 
\frac{1}{Z(\beta)} \int [ d \phi(x,\tau) ]  \prod_x  \delta( \phi(x, - \beta/2)- \phi'(x'))  \prod_x   \delta( \phi(x, \beta/2)- \phi''(x''))  \, e^{- S_E}
\label{cc7}
\eeq
where $S_E = \int_{- \beta/2}^{\beta/2}  d \tau \,  L_E$ with  $L_E$ the euclidean Lagrangian.  The normalization factor in (\ref{cc5}) guarantees that ${\rm Tr} \, \rho=1$.  The partition function $Z(\beta)$ is obtained doing the path integral with the identification $\phi'(x) = \phi''(x)$ at $\tau=- \beta/2$ and $\tau= \beta/2$, and integrating over these variables.  The geometry of the integration surface is a cylinder of length $\beta$. 

Let us now take a system $A$ given by the  interval $(u, v)$. One wants to compute the reduced density matrix $\rho(A)$ by tracing over the points not in $A$, that we call $B$. This operation amounts to gluing those points in (\ref{cc7}) and doing the integral over them. The effect is to leave a cut $(u, v)$ along the line $\tau= - \beta/2$. 
To compute ${\rm tr} \, \rho^n(A)$ one  uses the replica trick. One first makes $n$ copies of  (\ref{cc7}), labelled by $k=0, \dots, n-1$ and glue  them  together cyclically 

\beq
\phi'_k(x)  = \phi''_{k+1}(x), \quad (k=0, \dots, n-2), \qquad \phi'_{n-1}(x)  = \phi''_{0}(x), \qquad \forall  \, x \in A 
\label{cc8}
\eeq
\indent
The path integral on this $n$-sheeted geometry is denoted as $Z_n(A)$ and hence

\beq
{\rm tr} \, \rho^n(A)   = \frac{Z_n(A)}{Z(\beta)^n}
\label{cc9}
\eeq
\indent
CC argue that the LHS of (\ref{cc9}) is analytic for all $\Re \, n>1$ and that its derivative respect to $n$ in the limit
$n \rightarrow 1^+$ gives the entropy

\beq
S_A = - \lim_{n \rightarrow 1} \frac{ \partial}{ \partial n}  \frac{Z_n(A)}{Z(\beta)^n}
\label{cc10}
\eeq
\indent
Alternatively,  one can define the R\'enyi entropy $S_n(A)$ for the interval $A$, 

\beq
S_n(A) = \frac{1}{1-n} \log  {\rm tr} \, \rho^n(A) 
\label{cc11}
\eeq
and compute the entanglement entropy as

\beq
S_A = \lim_{n \rightarrow 1} S_n(A)  
\label{cc12}
.
\eeq
\indent
Let us now turn to the excited states in CFT that correspond to primary states. If $|0 \rangle$ is the vacuum state then  the incoming state generated by a primary operator  $\Upsilon(z, \bar{z})$, with conformal weights $h$ and $\bar{h}$,  is given by 

\beq
|\Upsilon \rangle = \lim_{z, \bar{z} \rightarrow 0} \Upsilon(z, \bar{z}) \, |0 \rangle
\label{cc13}
\eeq
while the outgoing state is given by

\beq
\langle \Upsilon | = \lim_{z, \bar{z} \rightarrow 0} \bar{z}^{- 2 h} \, z^{- 2 \bar{h}} \langle 0| \Upsilon^\dag \left(\frac{1}{\bar{z}}, \frac{1}{z}  \right) 
\label{cc14}
\eeq
where $\Upsilon^\dag$ is the operator conjugated to $\Upsilon$, i.e., $\Upsilon\times \Upsilon^\dag={\bf I}+\cdots$.
In the latter  two equations we have used the radial quantization so that $z = 0$ and  $z = \infty$ correspond to the infinite past and infinite  future respectively. The primary operator corresponding to the vacuum $|0 \rangle$ is, of course,  the identity, i.e. $\Upsilon = {\bf I}$. 
The density matrix of the ground state is given by the limit $\beta \rightarrow \infty$ of equation (\ref{cc7}), which we write as 

\beq
\langle  \{ \phi''(x'') \} | \rho_{\bf I} |  \{ \phi'(x') \} \rangle   = 
\frac{1}{Z} \int [ d \phi(x,\tau) ]  \prod_x  \delta( \phi(x, - \infty)- \phi'(x'))  \prod_x   \delta( \phi(x, \infty)- \phi''(x''))  \, e^{- S_E}
\label{cc15}
\eeq
where $Z= \lim_{\beta \rightarrow \infty}  Z(\beta)$ and the action $S_E$ is computed over a cylinder of infinite length. 
Based on equations (\ref{cc13}) and (\ref{cc14})  we find  the following expression for the density matrix of the primary state $|\Upsilon \rangle$, 

\beq
\langle  \{ \phi''(x'') \} | \rho_{\Upsilon} |  \{ \phi'(x') \} \rangle   = 
C   \int [ d \phi(x,\tau) ]  \prod_x  \delta( \phi(x, - \infty)- \phi'(x'))  \prod_x   \delta( \phi(x, \infty)- \phi''(x'')) \Upsilon(0, - \infty) \, \Upsilon^*(0, \infty)    \, e^{- S_E}
\label{cc16}
\eeq
which reduces to (\ref{cc15}) for  $\Upsilon = {\bf I}$. In this equation we have parametrized  the primary fields in terms of the space-time coordinates, i.e. $\Upsilon(\sigma, \tau)$, etc. not with  the radial coordinates, i.e. $\Upsilon(z, \bar{z})$, as in equations (\ref{cc13}) and (\ref{cc14}), but there is a one-to-one correspondences between the two. $C$ is a constant which will be fixed below.  Following the same steps as for the derivation of equation (\ref{cc9}), one finds that the trace of the reduced density matrix $\rho^n_\Upsilon(A)$ is given by

\beq
{\rm tr} \, \rho^n_\Upsilon(A)   = C^n Z_n(A) \langle \Upsilon_0(0, - \infty) \, \Upsilon^\dagger_0(0, \infty) \dots  \Upsilon_{n-1}(0, - \infty) \, 
\Upsilon^\dagger_{n-1}(0, \infty)  \rangle_{{\cal R}_n}  
\label{cc17}
\eeq
where ${\cal R}_n$ is the $n$-sheeted Riemann surface that results from the sewing of the $n$ copies of the original cylinder along the cuts associated to the interval $A$. The label $k =0, \dots, n-1$, of the primary field $\Upsilon_k$, and its conjugate $\Upsilon_k^\dagger$,  denotes the sheet they belong to. 
The  constant $C$ can now be fixed imposing the normalization of the reduced density matrix,

\beq
{\rm Tr} \, \rho_\Upsilon(A)   = C  Z  \langle \Upsilon_0(0, - \infty) \, \Upsilon^\dagger_0(0, \infty)   \rangle_{{\cal R}_1}   = 1
\label{cc18}
\eeq

Eliminating $C$ from this equation and plugging its  value into (\ref{cc17}) yields

\beq
{\rm tr} \, \rho^n_\Upsilon(A)   = \frac{Z_n(A)}{Z^n} 
\frac{  \langle \Upsilon_0(0, - \infty) \, \Upsilon^\dagger_0(0, \infty) \dots  \Upsilon_{n-1}(0, - \infty) \, \Upsilon^\dagger_{n-1}(0, \infty)  \rangle_{{\cal R}_n}  }{ 
 \langle \Upsilon_0(0, - \infty) \, \Upsilon^*_0(0, \infty)   \rangle_{{\cal R}_1}^n } 
\label{cc19}
\eeq
\indent
Finally, using equation (\ref{cc9}) in the limit $\beta \rightarrow \infty$, one gets 

\beq
F^{(n)}_\Upsilon(A)  \equiv \frac{ 
{\rm tr} \, \rho^n_\Upsilon(A)} {{\rm tr} \, \rho^n_{\bf I} (A)}   =  
\frac{  \langle \Upsilon_0(0, - \infty) \, \Upsilon^\dagger_0(0, \infty) \dots  \Upsilon_{n-1}(0, - \infty) \, \Upsilon^\dagger_{n-1}(0, \infty)  \rangle_{{\cal R}_n}  }{ 
 \langle \Upsilon_0(0, - \infty) \, \Upsilon^\dagger_0(0, \infty)   \rangle_{{\cal R}_1}^n } 
\label{cc20}
\eeq

Hence,  the ratio of the trace of the reduced density matrix of a primary state and that of the ground state,  is given essentially by a $2 n$ point correlator of the primary field $\Upsilon$ and its conjugate field $\Upsilon^\dagger$ on a $n-$sheeted Riemann surface. 
In the previous derivation we have not made used of the fact that $A$ is given by a single interval $(u,v)$, so equation (\ref{cc20}) also holds when $A$ consists in the union of multiple intervals. 

\begin{figure}[h]
\begin{center}
\includegraphics[height=3.75cm]{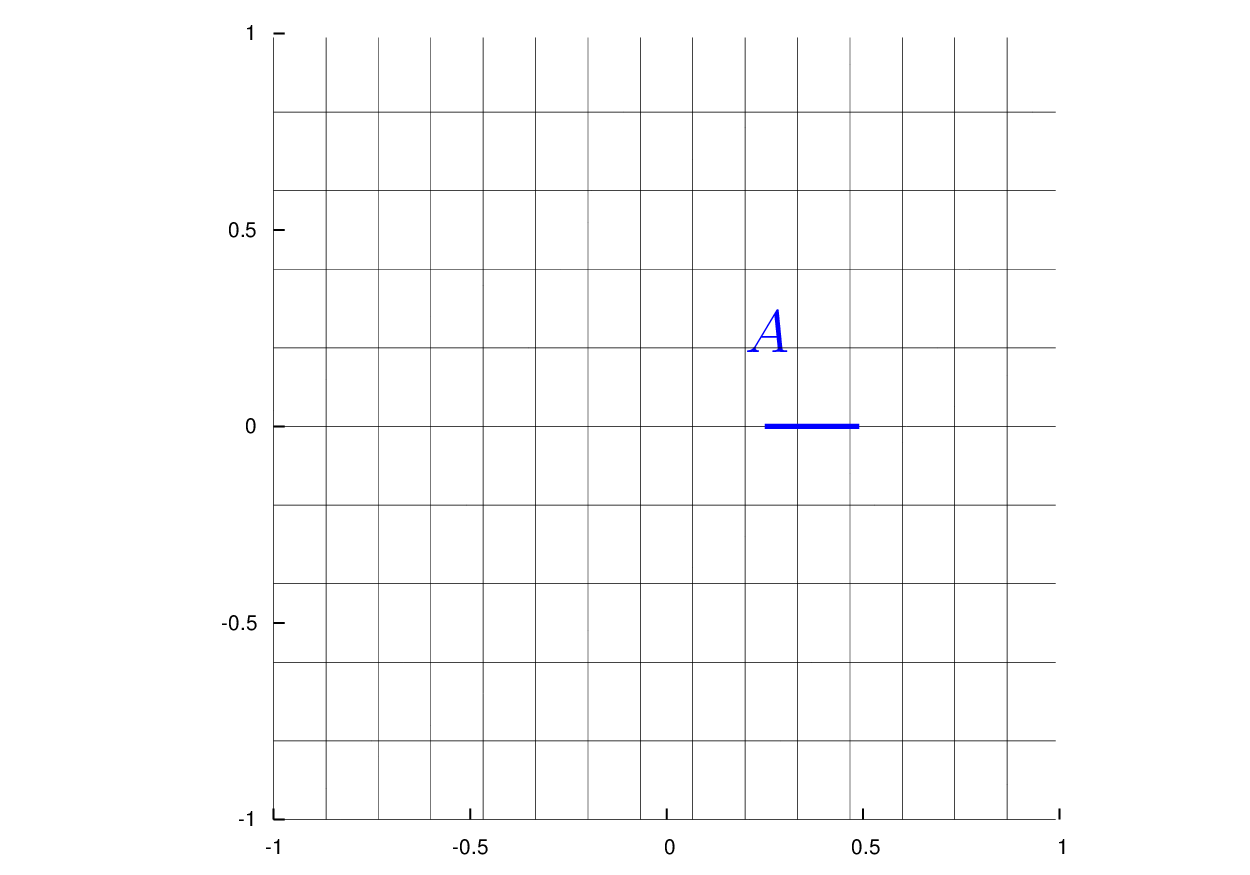}
\includegraphics[height=3.75cm]{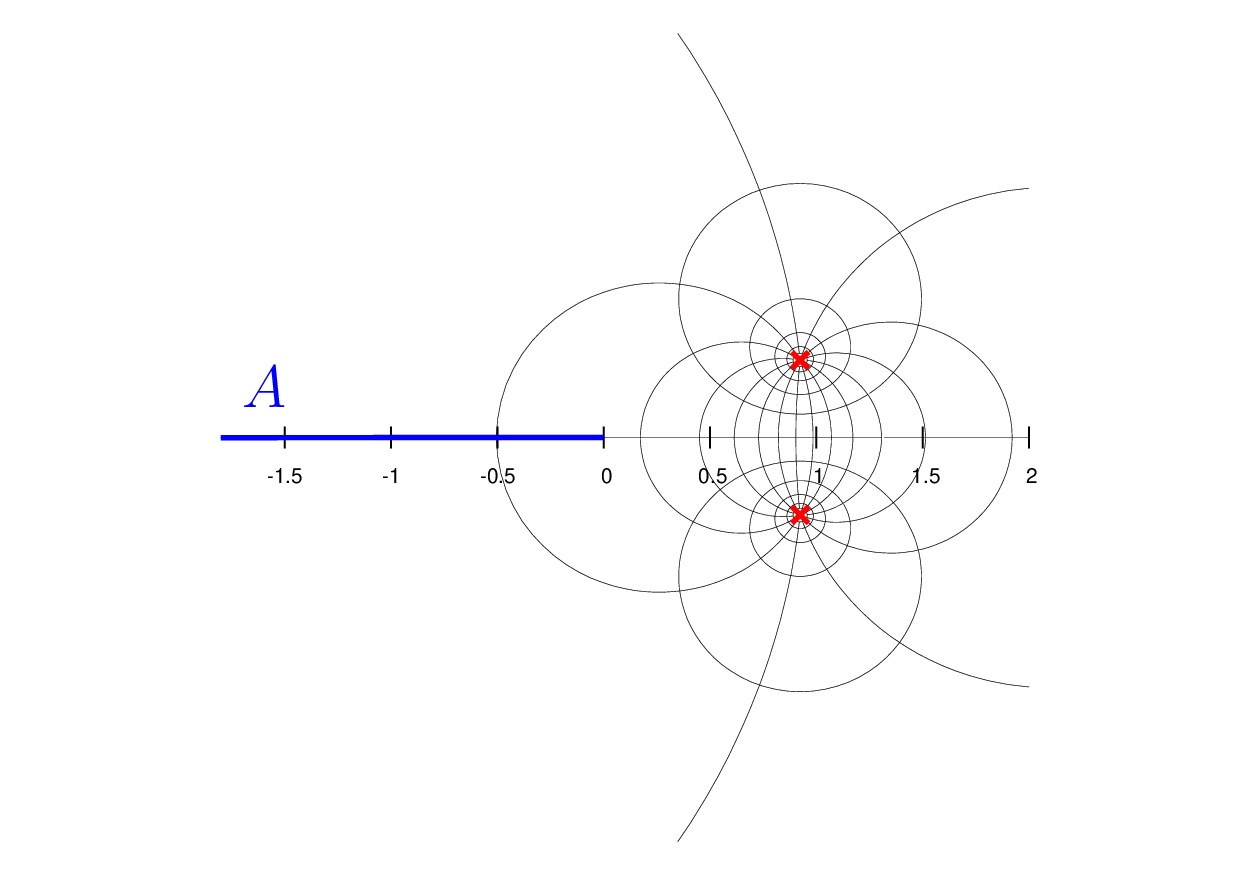}
\includegraphics[height=3.75cm]{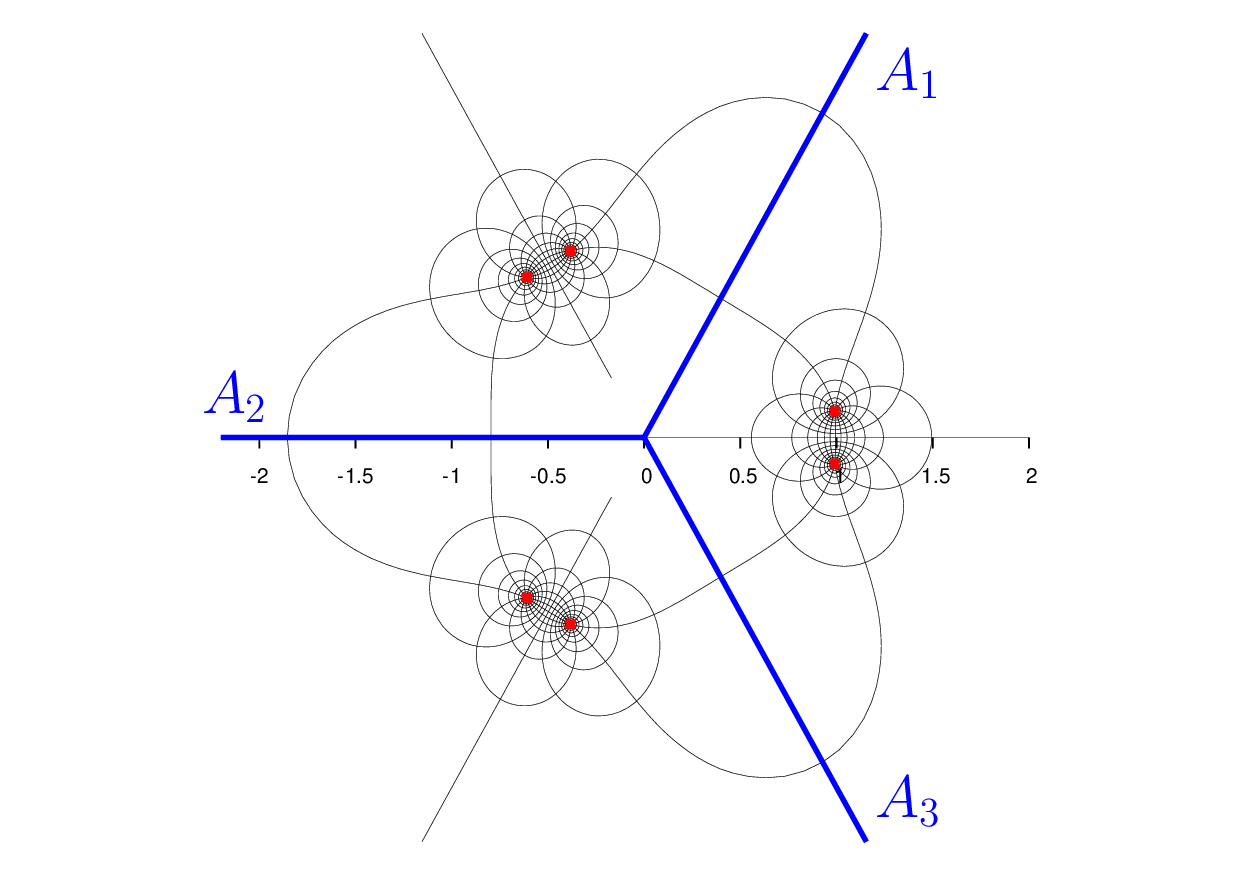}
\caption{\footnotesize The cylinder $w$ with the cut $A$ is mapped \textit{via} (\ref{cc21}) to the center panel (we took $x=|u-v|/L=1/8$). The transformation (\ref{cc24}) with $n=3$ has the effect shown in the right panel. The red crosses indicate the images $\zeta_\infty$, $\zeta'_\infty$, $z_{k,3}$, $z'_{k,3}$ of the points $w_\infty$, $w'_\infty$. }
\label{maps}
\end{center}
\end{figure}

Let us next compute the correlators appearing in equation (\ref{cc20}) in the case where $A= (u,v)$. To this end we shall first introduce the complex coordinate $w= \sigma + i \tau$, which parametrizes an  infinite cylinder of length $L$, i.e. $\sigma = \sigma + L$. The interval $A$ will be identified with the domain $u < w  < v$. This cylinder can be mapped into the complex plane by the conformal transformation

\beq
\zeta = \frac{ \sin \frac{ \pi ( w-u)}{L}}{ \sin \frac{ \pi (w-v)}{L}}
\label{cc21}
\eeq
\indent
The points   $w=u$ and $w=v$ are mapped into $\zeta=0$ and $\zeta = - \infty$ respectively,  so that the interval $A$ corresponds to the negative real axis $\Re \,  \zeta <0$. Moreover, the regions $\Im \,  w  \lessgtr 0$  are mapped into the regions $\Im \,  \zeta \gtrless 0$.  The points $(0, - \infty)$ and $(0, \infty)$, appearing in equation (\ref{cc20}) correspond to the coordinates $w_\infty = - i \infty$ and $w'_\infty = i \infty$, and,  by equation (\ref{cc21}),  to the $\zeta$-coordinates (see fig \ref{maps})

\beq
w_\infty = - i \infty  \Longrightarrow \zeta_\infty = e^{ i \pi x}, 
\qquad
w'_\infty =  i \infty  \Longrightarrow \zeta'_\infty = e^{- i \pi x}, \qquad x \equiv \frac{\ell}{L} , \quad \ell = v-u 
\label{cc22}
\eeq
\indent
Notice that $x$ belong to the interval $(0,1)$ since $\ell \in (0,L)$. 
The Riemann surface ${\cal R}_n$ is constructed out by gluing cyclically $n$ copies of the complex $\zeta$-plane along the cut $A$.
The uniforming parameter associated to  ${\cal R}_n$ is given by 

\beq
z = \zeta^{1/n} = \left( \frac{ \sin \frac{ \pi ( w-u)}{L}}{ \sin \frac{ \pi (w-v)}{L}} \right)^{ 1/n}
\label{cc23}
\eeq

Consequently, each of  the points $\zeta_\infty$ and $\zeta'_\infty$ give rises to $n$ points  in the Riemann surface ${\cal R}_n$ with coordinates

\beq
z_{k,n} = e^{ \frac{ i \pi}{n} ( x + 2  k)}, \qquad z'_{k,n} = e^{ \frac{ i \pi}{n} (- x + 2 k)}, \qquad k=0, \dots, n-1 
\label{cc24}
\eeq
\indent
To compute (\ref{cc20}) we first write it as
 
 \beq
F^{(n)}_\Upsilon(A)  \equiv F^{(n)}_\Upsilon(x) = \lim_{ w \rightarrow - i \infty}  
\frac{  \langle \prod_{k=0}^{n-1}  \Upsilon_k(w, \bar{w}) \, \Upsilon^\dagger_k(-w, - \bar{w})  \rangle_{{\cal R}_n}  }{ 
 \langle \Upsilon_0(w, \bar{w}) \, \Upsilon^\dagger_0(- w,- \bar{w})   \rangle_{{\cal R}_1}^n } 
\label{cc25}
\eeq
\indent
The $2 n$ point correlator on ${\cal R}_n$, appearing  in the numerator,  can be transformed into a correlator on the complex $z$-plane by means on the conformal transformation of the primary field $\Upsilon$ 

\beq
\Upsilon(w, \bar{w}) = \left( \frac{dz}{dw} \right)^h  \left( \frac{d \bar{z}}{d \bar{w}} \right)^{ \bar {h}}   \Upsilon(z, \bar{z})
\label{cc26}
\eeq
and a similar equation for $\Upsilon^\dagger$. From equation (\ref{cc23}) one finds

\beq
\frac{dz}{dw} = \frac{z}{n} \frac{ 4 \pi}{L} \sin( \pi x) \;   ( e^{ \frac{ i \pi}{L} ( w- u)} - e^{-  \frac{ i \pi}{L} ( w- u)} )^{-1} 
( e^{ \frac{ i \pi}{L} ( w- v)} - e^{-  \frac{ i \pi}{L} ( w- v)} )^{-1} 
\label{cc27}
\eeq 
which in the limit $w \rightarrow  \pm i \infty$ becomes

\barray
w \rightarrow - i \infty, \quad z & \rightarrow &  z_{k,n} \Longrightarrow  \frac{dz}{dw}  \rightarrow  \frac{z_{ k,n}}{n} \; \Lambda 
\label{cc28} \\
w \rightarrow  i \infty, \quad z & \rightarrow &  z'_{k,n} \Longrightarrow  \frac{dz}{dw}  \rightarrow  \frac{z'_{ k,n}}{n} \; \bar{\Lambda}   
\earray 
where

\beq
\Lambda = \frac{ 4 \pi}{L} \sin( \pi x) e^{ - 2 \pi |w|/L} e^{ i \pi (u+v)/L} 
\label{cc29}
\eeq

In this limit the conformal transformation (\ref{cc26}) yields 

\barray
w \rightarrow - i \infty, \quad z & \rightarrow &  z_{ k,n} \Longrightarrow  \Upsilon_k(w, \bar{w}) \rightarrow   \left( \frac{z_{ k,n}}{n} \; \Lambda \right)^h 
 \left( \frac{\bar{z}_{ k,n}}{n} \;  \bar{\Lambda} \right)^{\bar {h}}  \,  \Upsilon(z_{k,n}, \bar{z}_{k,n})
\label{cc30} \\
w \rightarrow  i \infty, \quad z & \rightarrow &  z'_{ k,n} \Longrightarrow \Upsilon_k^\dagger (w, \bar{w}) \rightarrow   \left( \frac{z'_{k,n}}{n} \; \Lambda \right)^h 
 \left( \frac{\bar{z}'_{k,n}}{n} \;  \bar{\Lambda} \right)^{\bar {h}}  \,  \Upsilon^\dagger(z_{k,n}, \bar{z}_{k,n})
 \earray 
Plugging these equations into (\ref{cc25}), the factors proportional to $\Lambda$ and $\bar{\Lambda}$ cancel out and one is left with 

\beq
 F^{(n)}_\Upsilon(x) = n^{ -2  n (h + \bar{h})}  \; 
 \frac{  \prod_{k=0}^{n-1}  z_{k,n}^h \bar{z}^{\bar {h}}_{k,n}  z_{k,n}'^h \bar{z}'^{\bar {h}}_{k,n}  }{
  \left(  z_{0,1}^h \bar{z}^{\bar {h}}_{0,1}  z_{0,1}'^h \bar{z}'^{\bar {h}}_{0,1} \right)^n} 
\frac{  \langle \prod_{k=0}^{n-1}  \Upsilon(z_{k,n}, \bar{z}_{k,n}) \, \Upsilon^\dagger(z'_{k,n}, \bar{z}'_{k,n})  \rangle_{{\mathbb C}}  }{ 
 \langle \Upsilon(z_{0,1}, \bar{z}_{0,1}) \, \Upsilon^\dagger(z'_{0,1}, \bar{z}'_{0,1})   \rangle_{{\mathbb C}}^n }
\label{cc31}
\eeq
\indent
To simplify this expression,  we can make a further conformal transformation from the complex plane to a cylinder of length $2 \pi$, 

\beq
z = e^{ i t} \rightarrow \Upsilon(t, \bar{t}) = e^{ i \pi ( h - \bar{h})} z^{h} \, \bar{z}^{\bar{h}} \; \Upsilon(z, \bar{z})
\label{cc32}
\eeq
which allow us to write (\ref{cc31}) as

\beq
 F^{(n)}_\Upsilon(x) = n^{ -2  n (h + \bar{h})}  \; 
\frac{  \langle \prod_{k=0}^{n-1}  \Upsilon(t_{k,n}, {t}_{k,n}) \, \Upsilon^\dagger(t'_{k,n}, {t}'_{k,n})  \rangle_{\rm cy}  }{ 
 \langle  \Upsilon(t_{0,1}, {t}_{0,1}) \, \Upsilon^\dagger(t'_{0,1}, {t}'_{0,1})   \rangle_{\rm cy}^n }
\label{cc31bis}
\eeq
where

\beq
t_{k,n} = \frac{\pi}{n} ( x + 2 k), \qquad t'_{k,n} = \frac{\pi}{n} ( -x + 2 k), \qquad k=0,1, \dots, n-1
\label{cc32}
\eeq
\indent
Since the latter parameters are all real we may as well write (\ref{cc31}) in a compact form as

\beq
 F^{(n)}_\Upsilon(x) = n^{ -2  n (h + \bar{h})}  \; 
\frac{  \langle \prod_{k=0}^{n-1}  \Upsilon( \frac{\pi}{n} ( x + 2 k) ) \, \Upsilon^\dagger( \frac{\pi}{n} ( -x + 2 k) )  \rangle_{\rm cy}  }{ 
 \langle  \Upsilon( \pi x ) \, \Upsilon^\dagger(- \pi x)   \rangle_{\rm cy}^n }
\label{cc33}
\eeq
where $\Upsilon(t) \equiv \Upsilon(t,t)$. After a shifting of the points in which the correlator is evaluated, this equation become (\ref{Fn}), which was derived in reference \cite{Alcaraz2011Entanglement} using the HLW approach. 
The function $ F^{(n)}_\Upsilon(x)$ satisfies the following identities

\beq
 F^{(n)}_\Upsilon(x)  =  \left(  F^{(n)}_\Upsilon(x) \right)^*= F^{(n)}_{\Upsilon^\dagger}(x) =  F^{(n)}_\Upsilon(1-x) 
\label{cc34}
\eeq
that can be derived from equation (\ref{cc31}) using the transformation properties of correlators under inversion $z \rightarrow - 1/z$ and rotations $ z \rightarrow e^{ i \alpha} z$.  The last equality in (\ref{cc34}) reflects the well know fact that the R\'enyi entropy $S_n(A)$ is the same as $S_n(B)$, where $B$ is the complement of $A$, which amounts to the replacement $ \ell \rightarrow L - \ell$, that is  $x \rightarrow 1-x$.

\section{Bosonic theory: $XX$, $XXZ$ and excluded-$XXZ$ models} \label{bosons}

The first theory under study is a massless bosonic field $\varphi(z,\bar z)$, with action:

\beq
\mathcal{A}[\varphi]=\frac{1}{8\pi}\int dz\, d\bar z\,\partial_z\varphi\,\partial_{\bar z} \varphi
\label{bosonaction}
\eeq
\indent
This is a CFT with central charge $c=1$ and two types of primary fields. The first type  is given by the vertex operators 
\beq
V_{\alpha,\bar \alpha}\equiv\, :e^{\myi (\alpha \phi+\bar \alpha\bar \phi)}:\,
,
\eeq
where $\phi$ , $\bar \phi$ are the chiral and anti-chiral parts of the bosonic field, solution of the equation of motion: $\varphi(z,\bar z)=\phi(z)+\bar \phi(\bar z)$. In the last equation $:\cdot:$ denotes normal ordering and $\alpha,\bar \alpha$ are real numbers. The vertex operator has conformal weights $(h,\bar h)$=$(\alpha^2/2,\bar \alpha^2/2)$ \cite{DiFrancesco1999Conformal}:

\beq
\<V_{\alpha_1,\bar\alpha_1}(z_1,\bar z_1)\,V_{\alpha_2,\bar\alpha_2}(z_2,\bar z_2)\> = (z_1-z_2)^{\alpha_1\alpha_2}\,(\bar z_1-\bar z_2)^{\bar \alpha_1\bar \alpha_2}
\eeq
where $\<\cdot\>$ denotes the correlator in the complex plane, and the \textit{neutrality condition} implies that the correlator is different from zero only if $\alpha_1+\alpha_2=0$ and the same for the $\bar \alpha$'s. This result is indeed more general. Considering holomorphic fields $\bar \alpha=0$ only, one has ($z_{ij}\equiv z_i-z_j$):

\beq
\<V_{\alpha_1}(z_1)\dots V_{\alpha_n}(z_n)\> = \prod_{i<j} [z_{ij}]^{\alpha_i\alpha_j}
\label{vertexcorrelators}
\eeq
if $\sum_j\alpha_j=0$, and zero otherwise. In a cylinder of length $2\pi$ parametrized by $w=-\myi \ln z$ this correlator is:

\beq
\<V_{\alpha_1}(w_1)\dots V_{\alpha_n}(w_n)\>_{\rm{cy}} = \prod_{i<j} [2\sin (w_{ij}/2)]^{\alpha_i\alpha_j}
\label{vertexcorrelatorscylinder}
\eeq
evaluating (\ref{vertexcorrelatorscylinder}) in the points indicated in (\ref{Fn}), i.e., in the set $\left\{2\pi j/n,2\pi(j+x)/n \right\}_j,\,j=0,\dots, n-1$, one has :

\beq
F^{(n)}_{V_{\alpha}}(x)=\left(n^{-n}\left[\frac{\sin(\pi x)}{\sin(\pi x/n)}\right]^n\,\,\prod_{m=1}^{n-1}\left[\frac{ \sin(\pi m/n)^2   }{ \sin{\frac{\pi(m-x)}{n}}\sin{\frac{\pi(m+x)}{n}}  }\right]^{m-n}\right)^{\alpha^2}
\label{Fnvertex}
.
\eeq
\indent
This function is constantly equal to one:

\beq
F^{(n)}_{V_{\alpha}}(x)=1 \qquad \forall n,\alpha
\label{Fnvertexis1}
,
\eeq
a fact that  can be proved by comparing the zeroes in $x$ of the numerator and in the denominator of (\ref{Fnvertex}). There are zeroes of order $n$ in $x\in\mathbb Z$ in both the numerator and the denominator. As analytic functions, this implies they are proportional, the proportionality constant being 1 as can be immediately seen in the $x\to 0$ limit. We thus observe that the entropy of excitations corresponding to vertex operators coincide with the ground state entropy according to the CFT prediction (\ref{Fn}).\\ \indent Let us now focus our attention in the other primary field in the theory: $\myi \partial\phi$, with conformal weights $(h,\bar h)=(1,0)$. We will use the correlators of $\myi \partial\phi$ in the plane:

\beq
-\<\partial\phi(z_1)\partial\phi(z_2)\>=\frac{1}{{z_{12}}^2}
\eeq
and the Wick theorem:

\beq
\<\prod_{j=0}^{2n} \myi \partial\phi(z_j)\>=\mbox{Hf}\,\left[\frac{1}{{z_{ij}}^2}\right]_{1\le i,j,\le 2n}
\label{2ncorrelators}
\eeq
where $\mbox{Hf\,[\,]}$ denotes the Haffnian ($S_{2n}$ is the permutation group of the $2n$ indexes):

\beq
\mbox{Hf}\,(A)=\frac{1}{2^n n!} \sum_{\pi\in S_{2n}} \prod_{i=1}^n\, A_{\pi(2i-1),\pi(2i)}
,
\eeq
and where the matrix in (\ref{2ncorrelators}) has a null diagonal. This matrix being dependent only on $i-j$, can actually be written as a determinant \cite{DiFrancesco1999Conformal}:

\beq
\mbox{Hf}\,\left[\frac{1}{{z_{ij}}^2}\right]=\mbox{det}\,\left[\frac{1}{z_{ij}}\right]
\label{Hfdet}
.
\eeq
\indent
In the cylinder $w=-\myi\ln z$, and taking again the $2n$ coordinates $w_j$ as in (\ref{Fn}) one gets:

\barray
F^{(n)}_{\myi\partial\phi}(x)=n^{-2n} \left[\sin{\pi x}\right]^{2n}\,\mbox{det}\,\left[\frac{1}{\sin{(w_{ij}/2)}}\right]_{1\le i,j\le 2n}
\label{FnUpsilon1}
,
\earray
\indent
The $n=2,3$ values of (\ref{FnUpsilon1}) are reported:

\barray
F^{(2)}_{\Upsilon_1} (x)  = 1 - 2 \mn r^2  + 3 \mn r^4 - 2 \mn r^6  + \mn r^8
\label{F2epsilon}
\earray
\barray
F^{(3)}_{\Upsilon_1} (x)  = \frac{1}{729}\left[1-4 \mn{s}  + \frac{2\mn{s}^3}{\mn{s}_+}-
\frac{2\mn{s}^3}{\mn{s}_-}+
\frac{3\mn{s}^2}{\mn{s}_+\mn{s}_-}-\frac{\mn{s}^3}{\mn{s}_+^3}+\frac{\mn{s}^3}{\mn{s}_-^3}\right]^2
\label{F3epsilon}
\earray
where $\mn r\equiv \sin  \pi x/2$, $\mn s\equiv \sin{\pi x/3}$, $\mn s_\pm \equiv \sin{\pi (1\pm x)/3}$. These values will be compared below with the numerical values of $F^{(n)}$ obtained for the  finite-size states $|\myi \partial\phi\>$ in the lattice models described by the bosonic theory.

\begin{figure}
\begin{center}
\includegraphics[height=7cm]{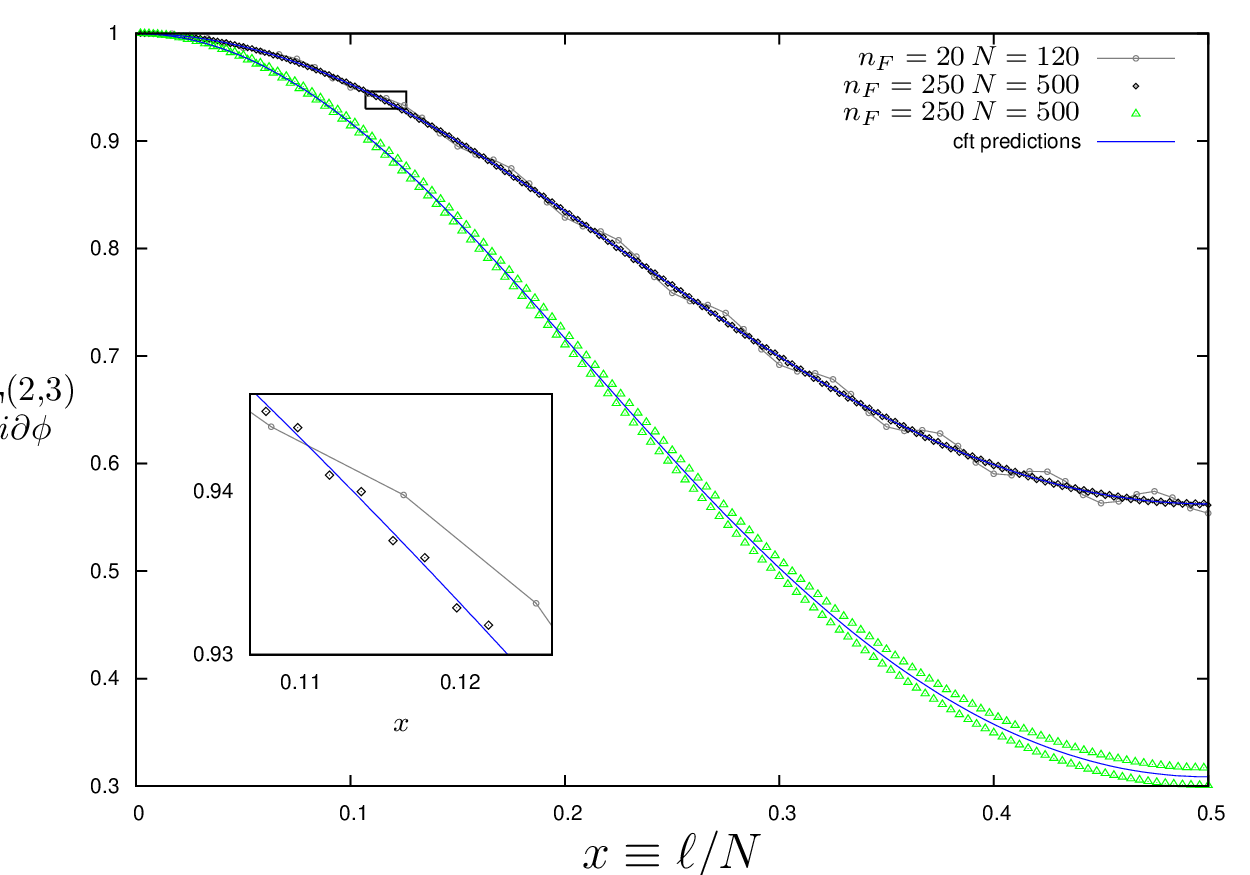}
\caption{\footnotesize R\'enyi entropy ratio of the particle-hole excitation in the $XX$/free fermion model, $n=2,3$, and two different sizes, vs. the CFT prediction equations (\ref{F2epsilon}, \ref{F3epsilon}) (continuous lines). For $n=2$, a system with filling fraction 1/6 is also shown. The inset is a zoom of the region defined by the small rectangle over the data-set in the body of the figure.}
\label{fermions-p1}
\end{center}
\end{figure}

\begin{figure}[h]
\begin{center}
\includegraphics[height=7cm]{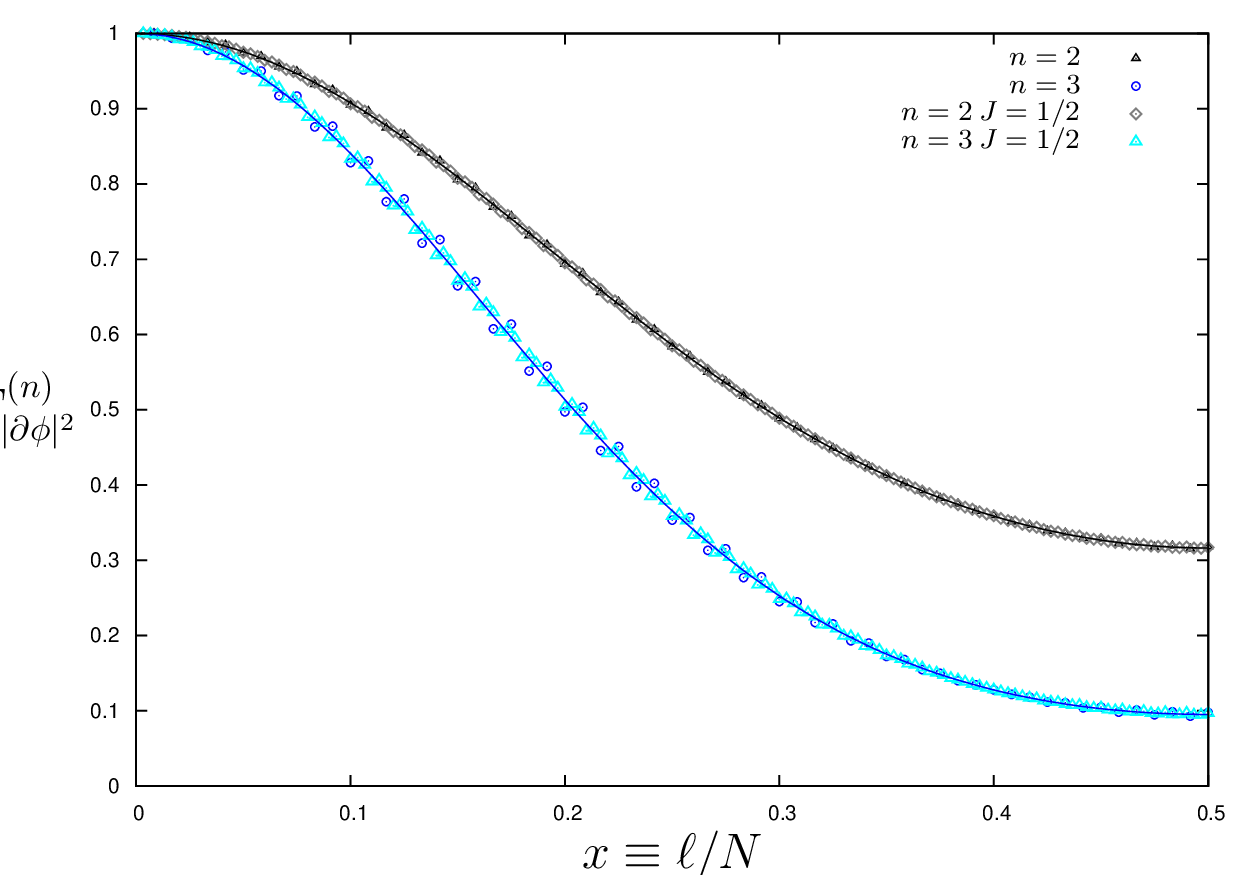}
\caption{\footnotesize $F^{(2,3)}_{|\partial \phi|^2}$ for a right-left particle-hole excitation in the $XX$ model with $N=120$ sites. Two values of the external field $J$ are shown: for $J=0$ and $J=1/2$ the filling fractions are $1/2$ and $1/3$ respectively. Continuous lines (indistinguishable from the numerical data for $n=2$) are the CFT prediction.}
\label{fermions-phRL}
\end{center}
\end{figure}

\subsection{$XX$ model}

As explained in the appendix, the $XX$ model is mapped into the lattice fermion problem (\ref{Hfermions}). In the continuum limit, this becomes a massless Dirac fermion whose low-energy regime is exactly formulated, via the bosonization procedure, as a free bosonic theory (\ref{bosonaction}). According to this map, $e^{\myi \phi}$ ($e^{-\myi \phi}$) corresponds in the fermionic language to the creation of a fermion (hole) (c.f. table \ref{mytable}).\\
\indent
Let us illustrate the law $F^{(n)}_{V_{\alpha,\bar \alpha}}(x)=1$ for three different states of the finite-size $XX$ problem. Consider $|V_{\alpha,\bar \alpha}\>$ with $(\alpha,\bar\alpha)=$ (1,-1). It is the \textit{Umklapp} excitation (c.f. table \ref{mytable}) which, as said in section \ref{XXoverview}, presents a vanishing excess of entropy. The same applies for the state  $(\alpha,\bar \alpha)=$ (-1,0), which corresponds to the $(a)$ excitation; to $(\alpha,\bar \alpha)=$ (1,1) which corresponds to the $(b)$ state (c.f. figure \ref{fermions-1stexcs}), and, in general, to all the low-energy compact excitations.  \\
\indent
We now focus in the state $|\myi \partial \phi\>$ with total momentum $P=2\pi/N$. In the $XX$/free fermion model it corresponds to the particle-hole excitation (c.f. table \ref{mytable}). In figure \ref{fermions-p1} we present the quantity $F^{(2,3)}_{\myi \partial \phi}$ for lattices with different sites and filling fractions, together with the CFT predictions (\ref{F2epsilon}), (\ref{F3epsilon}). A further prediction is shown in figure \ref{fermions-phRL} for the excitation $\partial\phi\bar \partial\bar\phi$, which corresponds in the lattice to a particle-hole excitation on the top of both the right and left Fermi points (c.f. table \ref{mytable}). The quantity $F^{(2,3)}_{\partial\phi\bar \partial\bar\phi}$ is compared with the CFT prediction, which is the square of (\ref{F2epsilon},\ref{F3epsilon}). We observe a very good agreement. \\
\indent
Finally, in reference \cite{Alcaraz2011Entanglement} it was shown how,  in the $x<<1$, $n\to 1$ limits,  (\ref{Fn}) results in the law (\ref{lowxS}), which was tested in the $XX$ model in figure  \ref{fermions-lowx}. Interestingly enough, and as can be seen in the figure, the law (\ref{lowxS}) works also for non-primary states.

\subsection{$XXZ$ and excluded-$XX$ models}

\begin{figure}[h]
\begin{center}
\begin{tabular}{cc}
\includegraphics[height=6cm]{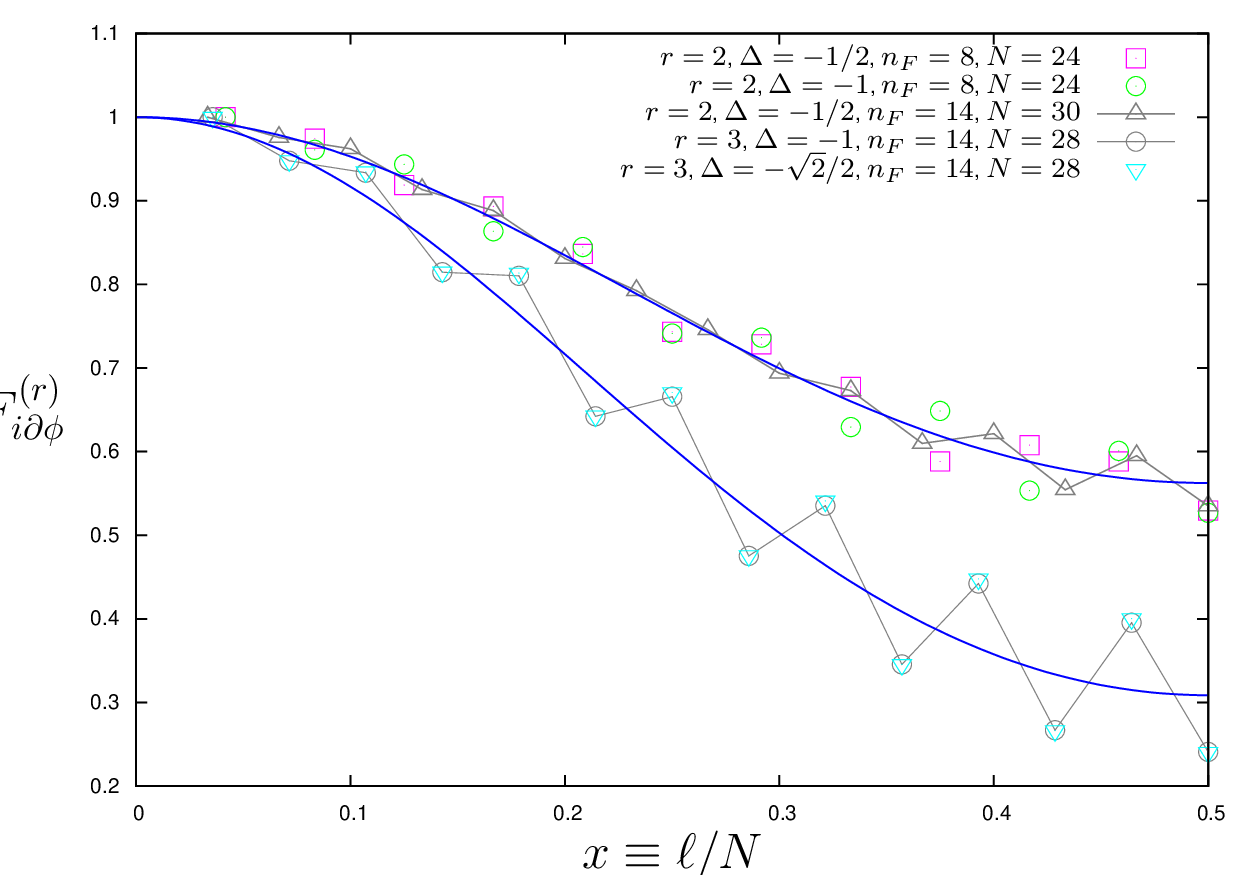} &
\includegraphics[height=6cm]{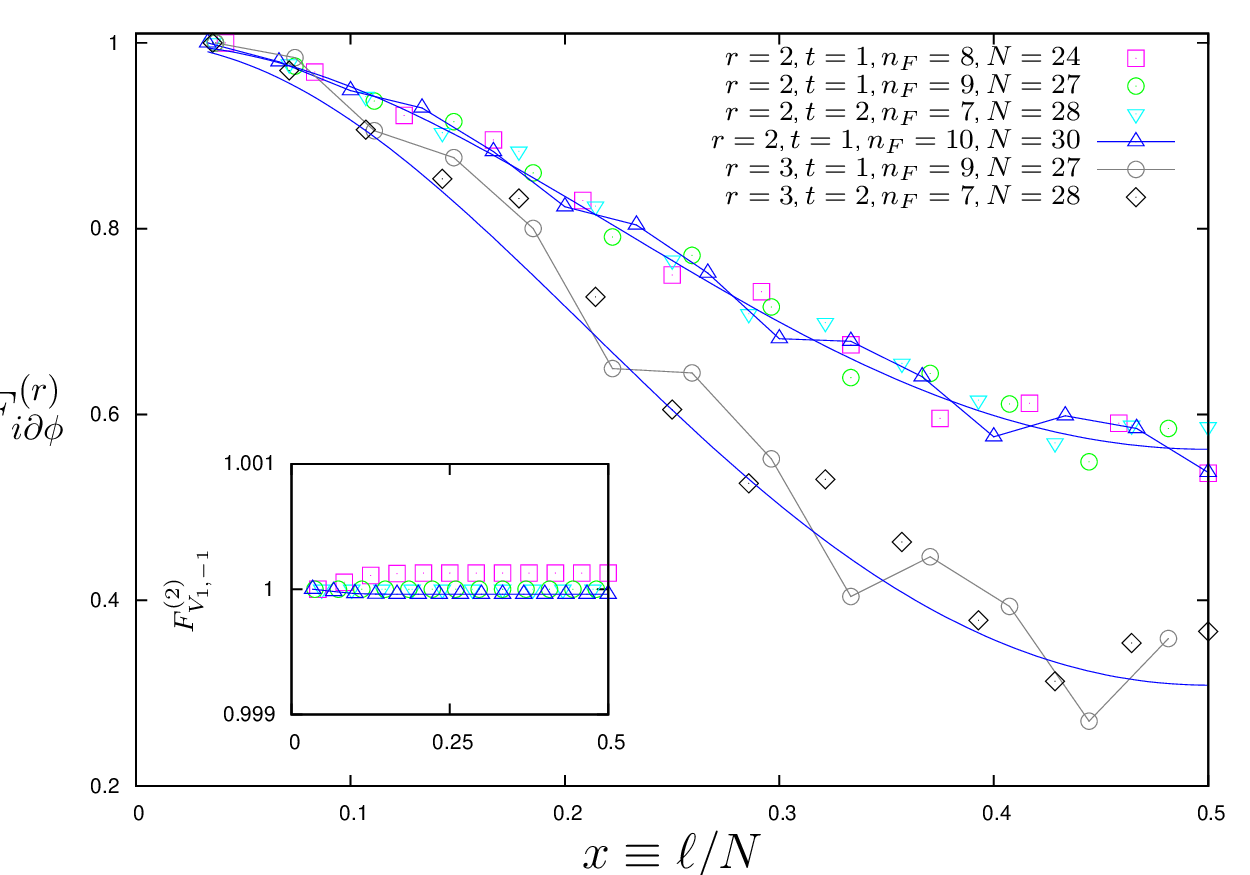}\\
(a) & (b)
\end{tabular}
\caption{\footnotesize Entropy ratio $F^{(r)}_{\myi \partial\phi}$, $r=2,3$, for the particle-hole state of the $XXZ$ model for different sizes and values of the anisotropy $\Delta$ in the critical region (left) and for the excluded-$XX$ model for $t=1,2$ and several filling fractions (right). On inset the Umklapp excitation is studied instead, which confirms the law (\ref{Fnvertexis1}).}
\label{xxz-oscn}
\end{center}
\end{figure}

\indent
We define the $XXZ$ spin chain:

\beq
H_{ XXZ}=-\frac{1}{2}\sum_{j=1}^N\left[ \sigma_j^x\sigma_{j+1}^x + \sigma_j^y\sigma_{j+1}^y+\Delta \sigma_j^z\sigma_{j+1}^z \right],
\label{XXZ}
\eeq 
where PBCs are assumed (for $\Delta=0$ one gets the $XX$ model). This model is integrable \cite{Yang1966OneDimensional} and gapless for $-1\leq \Delta <1$. In the continuum limit it is described  by the aforementioned bosonic CFT compactified in a circle of radius $K=\pi/(2\cos^{-1}(\Delta))$.  We had access to the exact entanglement entropy of this model through its exact diagonalization. \\
\indent
We first consider  the vertex states $|V_{\alpha}\>$ in the $XXZ$ model. The studied excitations were the following: the $(a)$ and $(b)$-types defined in table \ref{mytable}, the Umklapp state and the states $|V_\alpha\>$ with $\alpha=1,2,3$ (c.f. \cite{Alcaraz2011Entanglement}). We had access to the exact wave-function of the desired states (and hence to the entropy) through exact diagonalization. We verified that, as predicted by CFT, the $n$-R\'eny entropy coincides (up to oscillations) with the ground state entropy for $n=2,3$, in systems with several critical values of $\Delta$, filling fractions and sizes up to $N=30$ sites.   \\
\indent
Secondly, we studied the state $|\myi \partial\phi\>$ which is the lowest energy one in the sector with total momentum $P=2\pi/N$. The function $F^{(2,3)}$ is compared with the CFT prediction in figure \ref{xxz-oscn}-(a).\\

The excluded-$XXZ$ model \cite{Alcaraz1999Exact,Alcaraz1999Exactly} is an exactly integrable extension of the standard $XXZ$ chain.  In this model the  up spins ($\sigma^z$-basis)  have an effective size $t+1$ ($t=0,1,2,\ldots$) in units of lattice spacing, which means that up spins are not allowed at distances smaller than $t+1$ sites. The model is defined by the Hamiltonian

\begin{equation} \label{excl}
H = -\frac{1}{2}\sum_{j=1} P_t ( \sigma_j^z\sigma_{j+1}^z + \sigma_j^y\sigma_{j+1}^y +
\Delta\sigma_j^z\sigma_{j+1}^z)P_t,
\end{equation}
where $\Delta$ is the anisotropy and $P_t$ projects out states in which pairs of up spins are at distance smaller than $t+1$ sites. For $t=0$ one recovers the standard $XXZ$ model (\ref{XXZ}) where we only exclude up spins at the same site.  Like the $XXZ$ model there is a $U(1)$ symmetry, due to the conservation of the number of up spins. The model is critical for $-1\leq \Delta < 1$, and is ruled by a Luttinger liquid CFT ($c=1$)  whose Luttinger parameter $K$ depends on the value of $t,\Delta$ and the density of up spins. For $\Delta=0$, i. e., the excluded $XX$ chain \cite{Alcaraz1999Exactly} $K$ is known analytically, namely $K = (1-t\rho)^2$ where $\rho$ is density of up spins ($0\leq \rho \leq \frac{1}{1+t}$).\\
\indent
For $t=1,2$ and $\Delta=0$ we find similar results as for the $XXZ$ model. Figure \ref{xxz-oscn}-(b) shows some numerical results for the vertex and the particle-hole excitation.


\subsection{Universality of R\'enyi entropy oscillations }

Specially important are the models described by Luttinger Liquid field theories, which are CFTs with central charge equal to one. Their ground states obey hence the law (\ref{HLWlaw}) for the R\'enyi entropy with $c=1$. On the other hand, finite-size ground states of these models were shown in reference \cite{Calabrese2010Parity} to obey (\ref{HLWlaw}) up to some oscillations whose amplitude turns out to be related to the Luttinger Liquid parameter which depends on the microscopic couplings of the Hamiltonian. In this subsection we investigate the oscillations of some elementary excitations of the $XX$ and $XXZ$ models.\\
\indent
For the ground state, the deviation with respect to the CFT prediction (\ref{HLWlaw}), $S_n^{\rm{CFT}}$, takes the universal form \cite{Calabrese2010Parity,Xavier2011Renyi,Cardy2010Unusual}:

\beq
S_n(\ell)-S_n^{\rm{CFT}}(\ell)=f_n\,\mathcal{F}_n(\ell/N)\,\cos(2 k_F \ell)\,|D(\ell,N) \sin k_F|^{p(n,K)}
\label{oscillations}
,
\eeq
where $k_F$ the Fermi momentum, $D(\ell,N)=\frac{N}{\pi}\sin{\pi\ell/N}$ is the \textit{chord distance}, $f_n$ is a non-universal constant and $\mathcal{F}_n$ is a universal function depending in general on the parity of $N$. The exponent $p(n,K)=-2K/n$ is a function of the R\'enyi index and of the Luttinger Liquid parameter, $K$. In the models studied in \cite{Calabrese2010Parity,Xavier2011Renyi,Cardy2010Unusual}, no oscillations are found for the von Neumann entropy (i.e. for $n=1$ in (\ref{oscillations})). We also observe this fact for all the excitations considered in this article. In the present section we will provide numerical evidence that the same behaviour (\ref{oscillations}) holds for low-energy excitations, the function $\mathcal{F}_n$ depends in a universal way on the given excitation and the exponent $p(n,K)$ is given by $-2K/n$, at least for some of the excitations considered.\\

\begin{figure}[h]
\begin{center}
\begin{tabular}{cc}
\includegraphics[height=6cm]{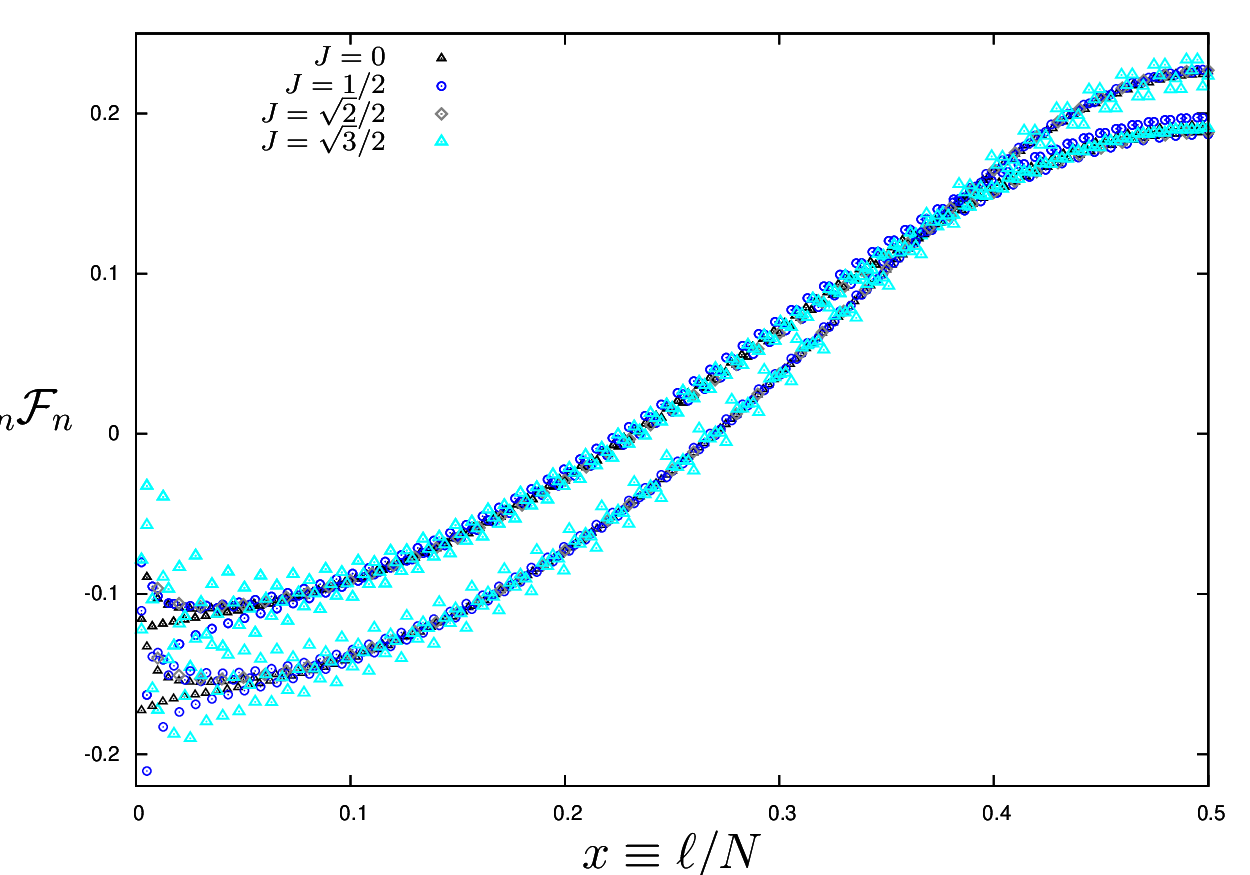} &
\includegraphics[height=6cm]{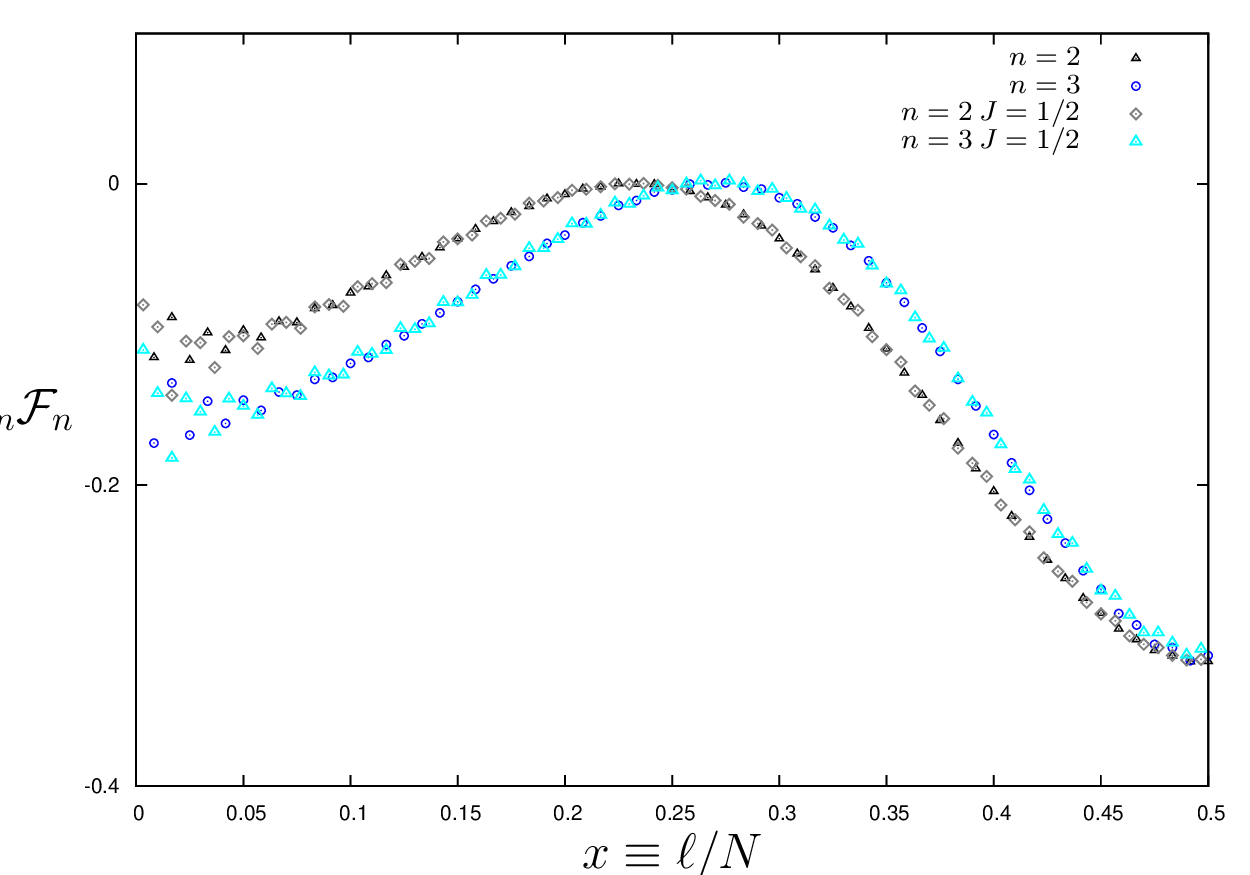} \\
(a) & (b)
\end{tabular}
\caption{\footnotesize (a) Function $f_n\,\mathcal{F}_n$ corresponding to the particle-hole excitation of the $XX$ model with four different values of the field $J$. $n=2,3$ cases are shown (the former being the upper curve at the half of the chain). (b) idem for the right-left particle-hole excitation. Colors are as in figure \ref{fermions-phRL}. }
\label{oscns-ph}
\end{center}
\end{figure}

For the $XX$ in an external $z$-field of strength $J$ (c.f. equation (\ref{Hxy}), $\gamma=0$), the law (\ref{oscillations}) holds for all the excitations considered. For this model $K=1$ regardless the value of $J$. The $N$-dependence of $S_n-S_n^{\rm{CFT}}$ at half of the chain shows a power law behaviour compatible with $p(n,1)=-2/n$ for $n=2,3$. Further numerical evidence can be found in figure \ref{oscns-ph} for the non-chiral and chiral particle-hole excitations: the same function $f_n\mathcal{F}_n$ can be found for common excitations in systems with different values of $J$ and filling fractions, showing the validity of equation (\ref{oscillations}) for $K=1$, the function $\mathcal{F}_n$ depending only on the excitation, and with $p(n,1)=-2/n$ for $n=2,3$.\\

\begin{figure}[h]
\begin{center}
\begin{tabular}{cc}
\includegraphics[height=5cm]{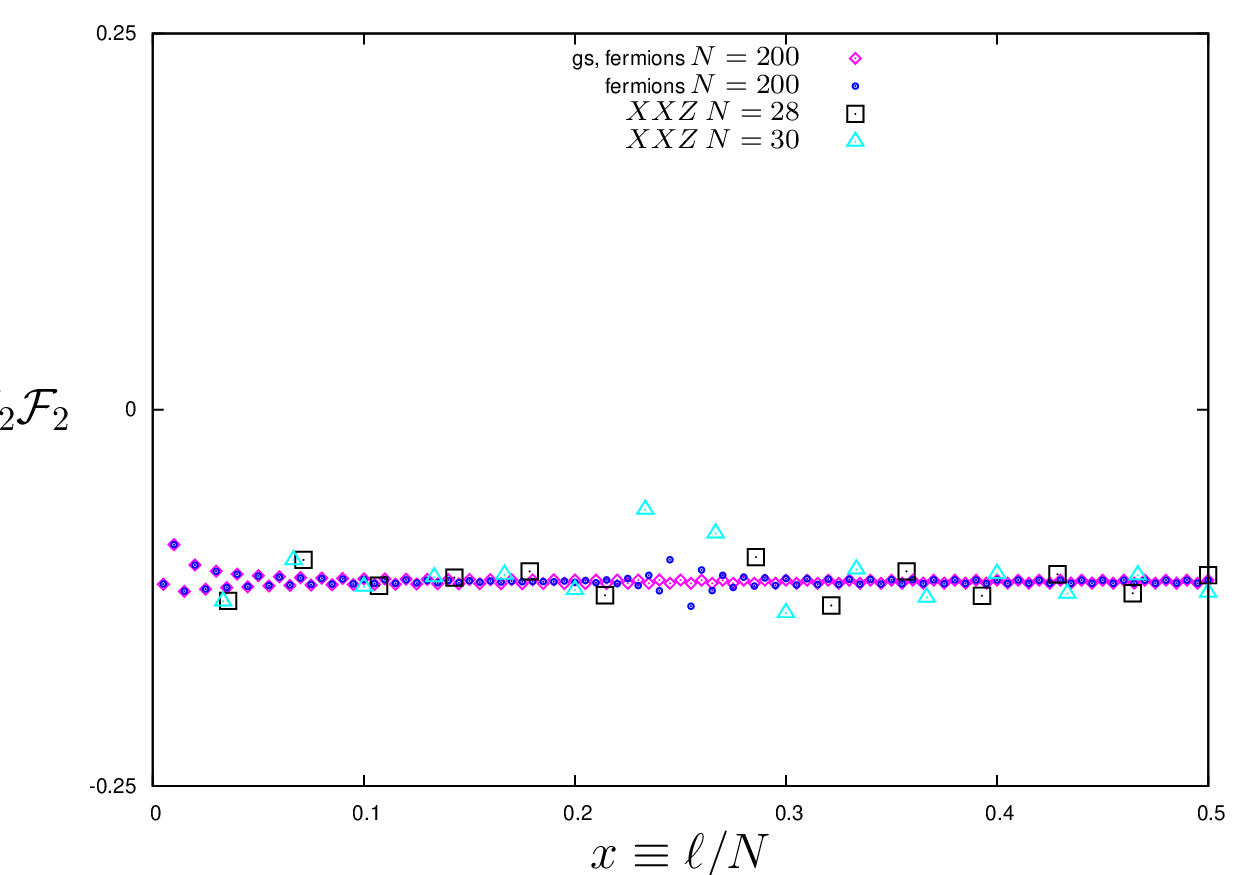} & 
\includegraphics[height=5cm]{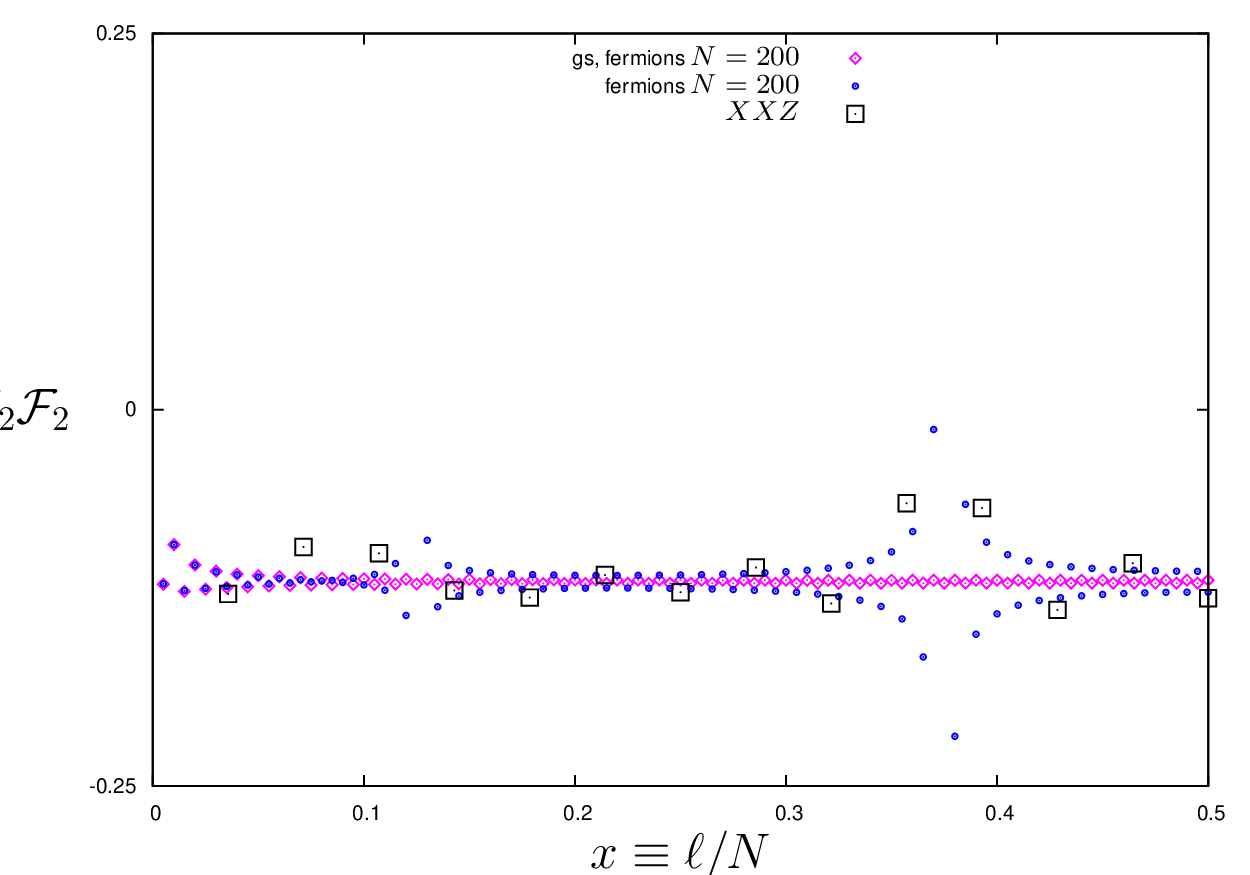}\\
(a)&(b)\\
\end{tabular}
\caption{\footnotesize Functions $f_n\,\mathcal{F}_n$ describing the amplitude of the R\'enyi oscillations for several excitations in the $XX$ and $XXZ$ ($\Delta=-1/2$) models with $N=240,500$ and $28$ respectively. (a) and (b) correspond to the ($a$) and ($b$) excitations defined in table \ref{mytable}. The universality of $\mathcal{F}_n$ is apparent, being this function equal with the ground state case, which is shown for comparison. }
\label{oscns}
\end{center}
\end{figure}

On the other hand, for vertex excitations one observes the validity of the scaling (\ref{oscillations}) with $p(n,K)$ also for $K\ne 1$. A first illustration of this fact can be found in the Umklapp excitation: as mentioned before, the state $|V_{1,-1}\>$ of the $XX$ model has exactly the same R\'enyi entropy as the ground state in the lattice. Moreover, the same  state in the $XXZ$ model exhibits the same entropy of the ground state up to corrections of the order of $10^{-3}$ (i.e., the same oscillations) for systems with 30 sites \cite{Alcaraz2011Entanglement} (see also figure \ref{xxz-oscn}-(a)). This implies that, for both $XX$ and $XXZ$ models, $\mathcal{F}_n$ is common to the ground state and the $|V_{1,-1}\>$ state. Further numerical evidence is shown in figure \ref{oscns}-a,b, where the ($a$) and ($b$) states in the $XX$ case are compared with the $XXZ$ case with $\Delta=-1/2$. The function $\mathcal{F}_n$  for vertex excitations turns out to be the same as for the ground state (apparently a constant function, c.f. figure \ref{oscns}, in agreement with \cite{Calabrese2010Parity}). We obtain analog results for other values of $\Delta$ and for $n=3$, with $p(n,K)=-2K/n$ as in (\ref{oscillations}). The case of the particle-hole excitation for $\Delta\ne 0$ is more controversial up to the sizes we have investigated, and will be studied in a future work which is in preparation \cite{IbanezBerganza2011Exactly}.

\section{Fermionic theory and entanglement in the critical Ising model} \label{Ising}

The second CFT we examine is the critical Ising model, whose action is given by

\beq
\mathcal{A}[\psi]=\frac{1}{8\pi}\,\int dz\, d\bar z\,\left[\psi\partial_{\bar z}\psi +\bar \psi \partial_{z} \bar \psi\right]
\label{fermionaction}
\eeq
where $\psi$ is a free Majorana fermion. The solution to the equations of motion being $\psi(z,\bar z)=\psi(z)+\bar \psi(\bar z)$. This theory is conformal, with $c=1/2$ and with primary operators $\sigma$ and $\psi$ having the conformal weights:

\beq
(h,\bar h)_\sigma=(\frac{1}{16},\frac{1}{16}),\qquad (h,\bar h)_\psi=(\frac{1}{2},0)
.
\eeq


Let us first study the function $F^{(n)}_\sigma$. One can compute \cite{DiFrancesco1999Conformal} the correlators of the field $\sigma$ by bosonization. Two Majorana fermions (labelled 1 and 2) can be combined into a Dirac Fermion $\Psi(z)=2^{-1/2}(\psi_1(z)+i\psi_2(z))$. $\Psi$ is in this way a (1/2,0) primary field with two-point correlator:

\beq
 \<\Psi(z_1)^\dag \Psi(z_2)\>= \<\psi(z_1)\psi(z_2)\>=\frac{1}{z_{12}}
\label{psicorrelator}
.
\eeq
\indent
The CFT describing the Dirac fermion has as central charge twice the Majorana fermion central charge, i.e., $c=1$. The Dirac theory can be mapped into a free bosonic theory with action (\ref{bosonaction}) in such a way that the spectrum of both theories coincide and that to each operator in the Dirac theory there corresponds an operator in the bosonic theory with common algebra and correlators. For the field $\Psi$, this mapping is:

\beq
\Psi(z)=e^{i\phi(z)},\qquad \bar \Psi(\bar z)=e^{i\bar \phi(\bar z)}
\nonumber 
\eeq
and for $\sigma$ (as can be checked from the matching of the conformal weights):

\beq
\sigma_1(z,\bar z)\sigma_2(z,\bar z)=2^{1/2} \cos (\varphi(z,\bar z)/2)
\nonumber 
.
\eeq
\indent
As described in \cite{DiFrancesco1999Conformal}, the following trick is used for the computation of a chain of $\sigma$ correlators. Two copies of $\sigma$ operator chains are constructed

\beq
\<\sigma_1(z,\bar z)\sigma_2(z,\bar z)\sigma_1(w,\bar w)\sigma_2(w,\bar w)\>=\<\sigma(z,\bar z)\sigma(w,\bar w)\>^2=2\left<\cos \frac{\varphi(z,\bar z)}{2}\,\cos \frac{\varphi(w,\bar w)}{2}\right>
\nonumber 
.
\eeq
\indent
Using the vertex operator correlator (\ref{vertexcorrelators}) one arrives to:

\beq
\<\sigma(z,\bar z)\sigma(w,\bar w)\>^2=|z-w|^{-1/2}
\nonumber
,
\eeq
and, for a  set of $2n$ fields, to:

\beq
\<\sigma(z_1\bar z_1)\dots \sigma(z_{2n}\bar z_{2n})\>^2=2^{-n}\sum_{\{\alpha_j=\pm 1\}}\, \prod_{i<j} |z_i-z_j|^{\alpha_i\alpha_j/2}
\nonumber
\eeq
where the sum is subject to: $\sum_j \alpha_j=0$. In the cylinder $w=-\myi \ln z$  this is:

\beq
\<\sigma(w_1\bar w_1)\dots \sigma(w_{2n}\bar w_{2n})\>^2=2^{-n}\sum_{\{\alpha_j=\pm 1\}}\, \prod_{i<j} |\sin w_{ij}|^{\alpha_i\alpha_j/2}
\eeq
\indent
With this information one can construct the function $F^{(n)}_\sigma$, which reads:\\

\barray
\left[F^{(n)}_\sigma(x)\right]^2=|\sin(\pi\, x)|^{n/2}
\sum_{\{\alpha_j=\pm 1\}} \left[\prod_{q=0}^{n-1}\,\prod_{j=1}^{n-q}\left|\sin \frac{\pi(q+x)}{n}\right|^{\frac{\alpha_{2j-1}\alpha_{2(j+q)}}{2}}\right.\nonumber \\
\left.\prod_{q=1}^{n-1}\,\prod_{j=1}^{n-q}\left|\sin \frac{\pi(q-x)}{n}\right|^{\frac{\alpha_{2j}\alpha_{2(j+q)-1}}{2}} \left|\sin \frac{\pi q}{n}\right|^{\frac{\alpha_{2j}\alpha_{2(j+q)}}{2}} \right]\,=1
\label{Fnsigma}
.
\earray
\indent
This function can be proved to be constantly equal to one for all values of $n$, in the same way that this fact was proved for $F^{(n)}_{V_\alpha}$. The square of equation (\ref{Fnsigma}) can be written as the product of two terms: the function $|\sin(\pi\,x)|^n$ on the one hand, which presents a zero of order $n$ for each $x\in\mathbb Z$, and the sum on the permutations squared, which presents poles in $x \in \mathbb Z$ of order $n$. Indeed, for each integer whose rest of the division by $n$ is $p>0$, there is a pole of order $n-p$ in the $q=p$ product in the second line of (\ref{Fnsigma}), and a pole of order $p$ in the $q=n-p$ product in the first line, and this happens since, for a given $p$, the permutation $\{\alpha_j=(-1)^j\}$ satisfies $\sum_{j=1}^{n-p}\alpha_{2j}\alpha_{2(j+p)-1}=n-p$ and $\sum_{j=1}^{p}\alpha_{2j-1}\alpha_{2(j+n-p)}=p$.  We  hence find the CFT prediction that the entropy of $\sigma$ excitations coincides again with the ground state entropy. \\
\indent
We repeat the game for the \textit{thermal field} $\varepsilon(z,\bar z) \equiv \myi \psi(z) \bar \psi (\bar z)$, with conformal weights $\left(h,\bar h\right)=\left(\frac{1}{2},\frac{1}{2}\right)$. For the computation of $F^{(n)}_\varepsilon$ we use the $2n$-point correlator of $\varepsilon$'s (cf. \cite{DiFrancesco1999Conformal}, p. 444), which can be inferred from the $\psi$ correlator (\ref{psicorrelator}): 

\barray
& \<\varepsilon(z_1,\bar z_1)\dots\varepsilon(z_{2n},\bar z_{2n})\>= \nonumber \\
&\<\psi(z_1)\dots\psi(z_{2n})\>\<\bar \psi(\bar z_1)\dots\bar \psi(\bar z_{2n})\>   = 
\left|\mbox{Pf}\,\left[\frac{1}{z_{ij}}\right]_{1\le i,j\le 2n}\right|^2
\label{epsilonstring}
,
\earray
where Pf means the Pfaffian:

\beq
\mbox{Pf}\,(A)=\frac{1}{2^n n!} \sum_{\pi\in S_{2n}} \mbox{sign}(\pi)\prod_{i=1}^n\, A_{\pi(2i-1),\pi(2i)}
.
\eeq
\indent
In the cylinder $w=-\myi\ln z$ and taking the $2n$ coordinates $w_{ij}$ as in (\ref{Fn}), one arrives to:

\barray
\<\prod_{j=1}^{2n} \varepsilon(w_j,\bar w_j)\> = \left( \mbox{Pf}\,\left[\frac{1}{\sin{(w_{ij}/2)}} \right]_{1\le i,j\le 2n}\right)^2
\earray
and using that the Pfaffian of an antisymmetric matrix is the square root of its determinant, one finds that $F^{(n)}_\varepsilon(x)$  coincides with the value computed in section \ref{bosons} for the field $\myi\partial\phi$ (c.f. equation (\ref{FnUpsilon1})):

\barray
F^{(n)}_\varepsilon(x)=F^{(n)}_{\myi\partial\phi}(x)
\label{Fnepsilon}
.
\earray
\indent
Finally,  using (\ref{epsilonstring}), one obtains:

\barray
F^{(n)}_\psi(x)=\sqrt{F^{(n)}_{\myi\partial\phi}(x)}
\label{Fnpsi}
.
\earray

\subsection{Ising model on the lattice}

Consider the critical Ising model on a transverse field (ITF)

\beq
H_I=-\frac{1}{2}\sum_{j=1}^N\,\left(\sigma_j^x\sigma_{j+1}^x+J\,\sigma_j^z\right)
.
\eeq
\indent
The model is integrable and critical for $J=1$. The critical model is described in   the continuum limit by the free Majorana fermion (\ref{fermionaction}) \cite{DiFrancesco1999Conformal}, as can be seen from the spectra of its free-fermionic formulation (\ref{Bogolubovspectrum}). We identified the states $|\sigma\>$, $|\varepsilon\>$, $|\psi\>$ in the finite size Ising model through their finite-size scaling of energy and momentum and for these states we computed the entanglement entropies and the quantity  $F^{(n)}$ \textit{via} the method described in \cite{Chung2001Densitymatrix} and in the appendix.\\
\begin{figure}[h]
\begin{center}
\includegraphics[height=7cm]{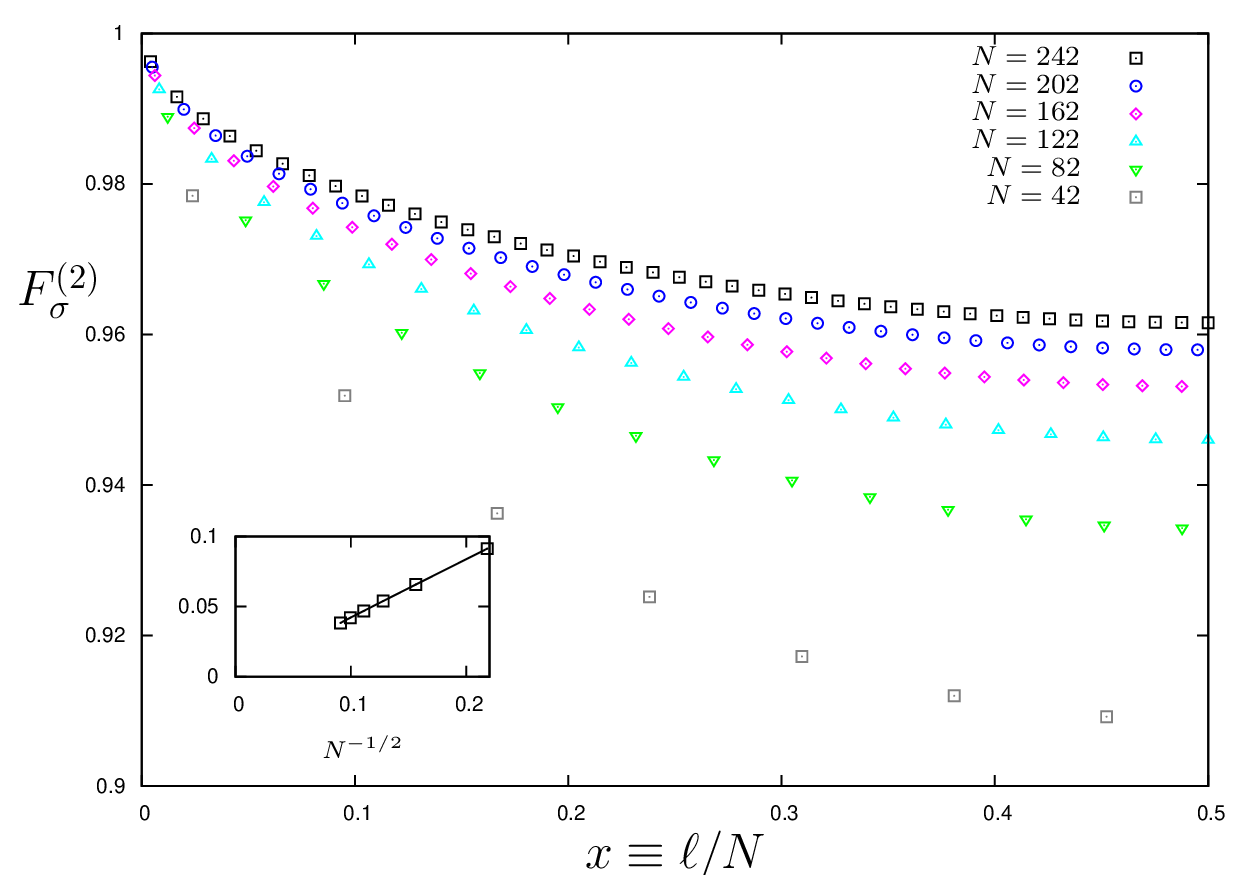}
\caption{\footnotesize The function $F^{(2)}_\sigma$ seems to converge to the CFT prediction $F^{(n)}_\sigma(x)=1$ in the large $N$-limit of the critical Ising model. The inset shows the $N^{-1/2}$-dependence of $F^{(2)}(1/2)$.}
\label{F2sigma}
\end{center}
\end{figure}

The excitation $|\sigma\>$ corresponds in the lattice to the lowest energy state in the parity -1 sector. Figure \ref{F2sigma}  shows the quantity $F^{(2)}_\sigma$  for systems of several sizes up to $N=242$. One observes that, although the $N$-scaling is slower than in the bosonic case, the curve seems to converge to the CFT prediction (\ref{Fnsigma}) for large $N$ (see $N$ scaling of $F_\sigma^{(2)}(1/2)$ in the inset). For $F^{(3)}_\sigma$ we obtain analog results.

\begin{figure}[h]
\begin{center}
\includegraphics[height=7cm]{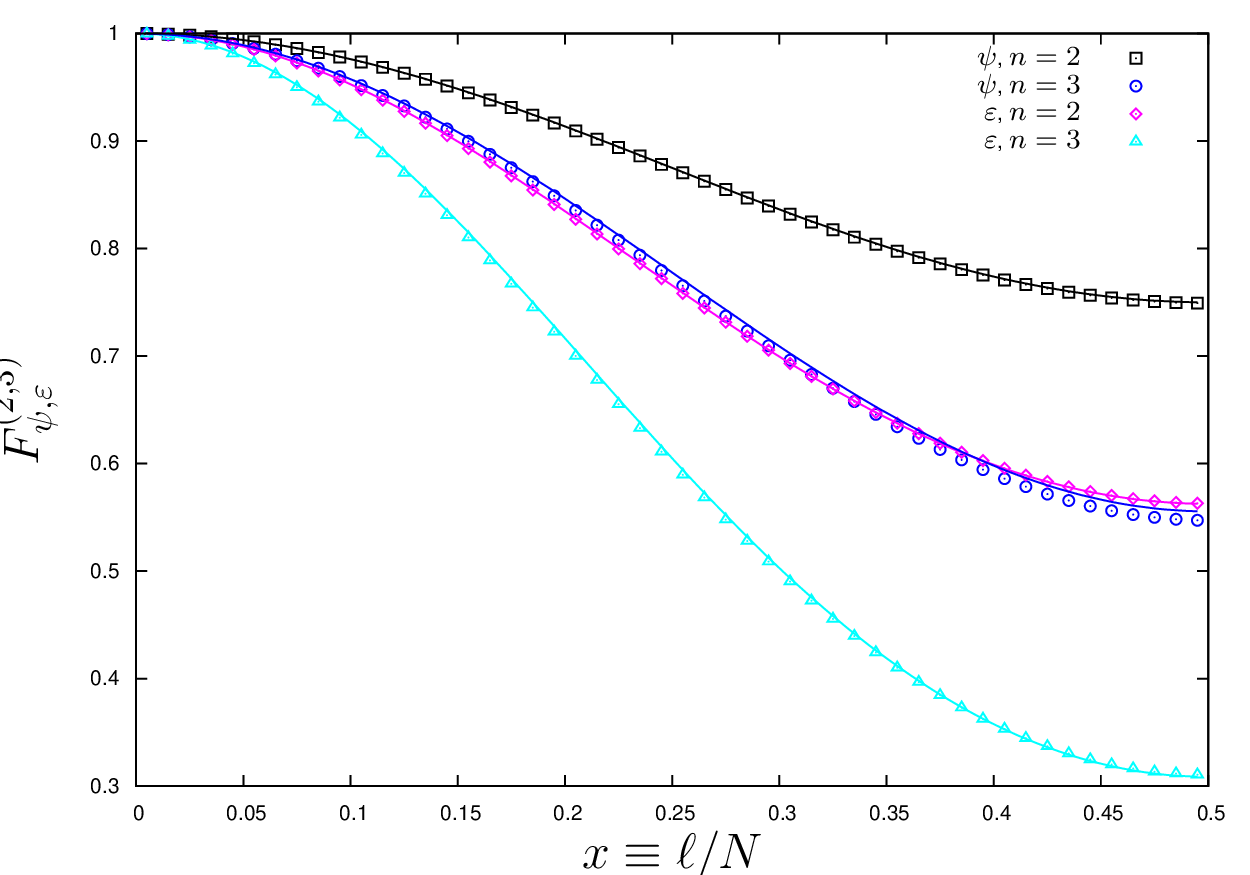}
\caption{\footnotesize $n=2$ and 3-R\'enyi entropy ratio for the states $|\psi\>$, $|\varepsilon\>$ of a $N=200$ critical Ising, compared with the CFT predictions for these fields (continuous curves).}
\label{F23epsilon}
\end{center}
\end{figure}

On the other hand, a comparison between the numerical values for the quantities $F^{(n)}_\psi$, $F^{(n)}_\varepsilon$, $n=2,3$ and the laws (\ref{Fnepsilon},\ref{Fnpsi}) is shown in figure \ref{F23epsilon}. Once more, the agreement between CFT and finite-site lattices is remarkable.

\section{Descendant fields} \label{descendant}

\begin{figure}[h]
\begin{center}
\includegraphics[height=7cm]{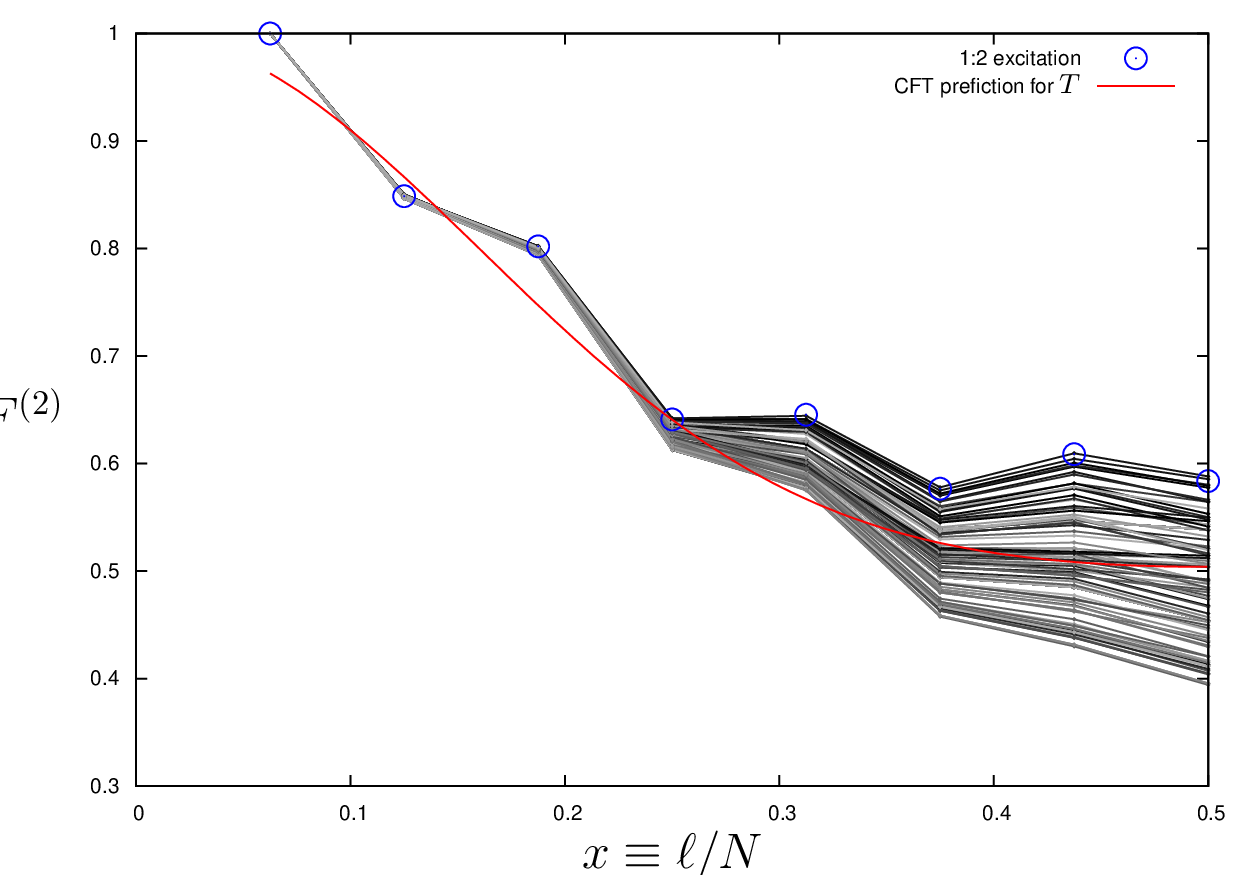}
\caption{\footnotesize $n=2$ entropy ratio of several states in the $(h,\bar h)=(2,0)$ sector of the $XX$ model with $N=16$.}
\label{S2deg}
\end{center}
\end{figure}

\begin{figure}[h]
\begin{center}
\includegraphics[height=7cm]{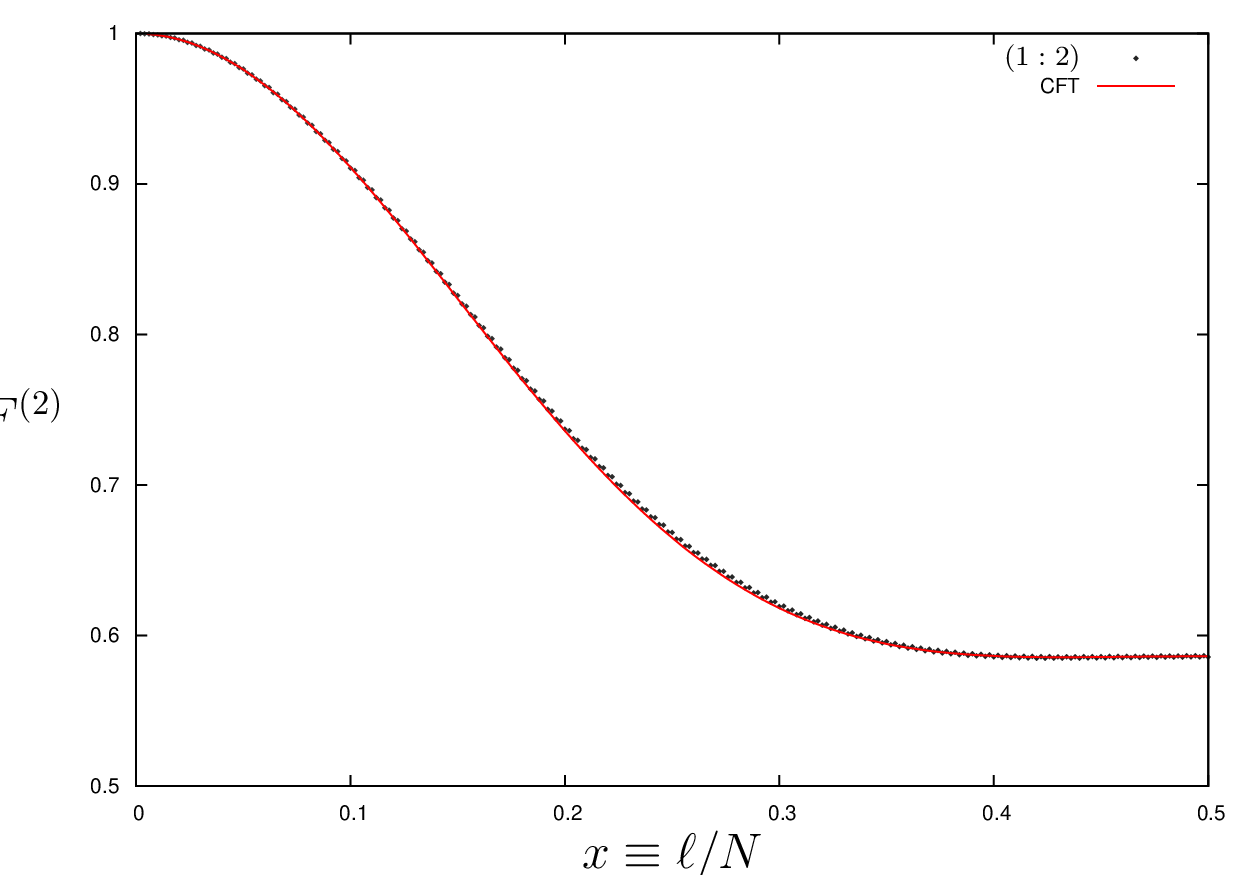}
\caption{\footnotesize $n=2$ entropy ratio corresponding to the $(1:2)$ excitation of the $XX$ model with $N=400$ (points). The continuous curve is the a fit with the function $F^{(2)}_Y$ with $Y$ given by (\ref{Yfield}).}
\label{F2deg}
\end{center}
\end{figure}

The results presented so far give a strong numerical  support  to  the validity of equation (\ref{cc33}) for   the entropy ratio $F^{(n)}_\Upsilon$ when $\Upsilon$ is a primary field.  A natural question is whether that  formula also applies when $\Upsilon$ is not a primary field but a descendant. We have not yet found a definite  answer to this question but there are some observations that may  eventually lead
to a solution (for details see \cite{IbanezBerganza2011Exactly}). Let us  first  recall that in  the field theory construction of the reduced density matrix,  $\rho_\Upsilon$, one   inserts  into  the path integral an  operator   at $\tau = - \infty$, and another one at  $\tau = \infty$, in order to  create the incoming and outgoing states associated to $\Upsilon$ and $\Upsilon^\dagger$. These operators are inserted in a cylindrical geometry and hence the corresponding operators  must be the conformal transformed of those defined on the complex plane. For primary fields,  the latter conformal transformation is rather simple but for descendant fields it generates additional fields  in the conformal tower. A typical example is the energy-momentum tensor, $T$, whose conformal transformation, from the complex plane to the cylinder,  involves a constant term proportional to the Schwarzian derivative. Hence in this case the operators to be inserted at $\tau = \pm \infty$ involve $T$ plus a constant term. However this constant term cancels out in the conformal transformation (\ref{cc23}) from the cylinder to the uniformizing plane $z$, so that the final expression for $F^{(n)}_T$ is given by equation (\ref{cc31}) with $\Upsilon$ replaced by $T$ in  the $2 n$-point correlator  on the complex plane. The same result can be obtained  for the descendant field $\partial^2 \phi$ of the $c=1$ theory, and we expect this result to hold in general. It would thus seem that  equation (\ref{cc31}) must  also be valid for descendant fields. However the numerical results give only a partial confirmation of this conjecture, which we now explain in more detail.\\
\indent
Let us first  consider the entropy of descendant fields in the sector with $(h,\bar h)=(2,0)$ of the bosonic CFT. This space is the first (two-dimensional) degenerated sector of the identity tower in the bosonic CFT, and  it is expanded by the fields $:(\partial\varphi)^2:$ and $:\partial^2\varphi:$ (the former being proportional to the stress-energy tensor of the theory, $T$). In the  $XX$ and $XXZ$  models this sector has  total energy and momentum $E=P=4\pi/N$.\\
\indent
In reference \cite{Alcaraz2011Entanglement} it was noticed that the entropy of degenerate sectors could depend on the particular state in the sector. We will provide a example of this fact. Let $|1\>$, $|2\>$ denote a basis of the $(h,\bar h)=(2,0)$ sector in a finite-size realization of the $XX$ chain. We computed the entropy $S_2$ for a general eigenvector of the $XX$ model in the mentioned sector:

\beq
|v\>=\cos\alpha |1\> + e^{i\beta}\sin \alpha |2\>
\eeq
\textit{via} exact diagonalization. Figure \ref{S2deg} shows  $F^{(2)}_{|v\>}$ for several values of $\alpha$ and $\beta$, together with the entropy of the state $(1:2)$ computed with the correlation matrix method exposed in the appendix (circles). The states $(1:2)$ and $(2:1)$ constitute a basis of the sector, an both have  the same entropy. Different grey levels in fig.\ref{S2deg}  denote different values of $\alpha$. There is a certain linear combination $|v\>$ for which the entropy coincides with the entropy of the state $(1:2)$, which turns out to have the lowest  value. The CFT prediction for $T$, obtained using equation (\ref{cc31}),  is also reported in figure \ref{S2deg} and it seems to coincide with one of the  states  $|v \>$.\\
\indent
As another example,  let us consider the  descendant  state $(1:2)$, whose field theory counterpart must be a linear combination $Y$  of the operators

\beq
Y=\cos\alpha\,\,\partial^2\phi + e^{i\beta}\sin \alpha\,\, T
\label{Yfield}
\eeq
\indent
The values of $\alpha$ and $\beta$ can be chosen  to fit the numerical values  of $F^{(2)}_{(1:2)}$ to the value of  $F^{(2)}_Y$ obtained using equation (\ref{cc31}). The result shown in fig \ref{F2deg}, shows a remarkable agreement, which however is not as good as in the case of primary fields. Another example is the entropy of the energy-momentum tensor of the Ising model, where the CFT prediction given by equation (\ref{cc31}) disagrees strongly with the numerical results. All this shows that the general application of this formula for descendants fields deserves further clarification.

\section{Conclusions and perspectives}

The main findings reported in this article are summarized in what follows. The universality of the ground state R\'enyi entropy of entanglement at criticality is generalized to low-energy excitations represented by primary fields in CFT, as stated in reference \cite{Alcaraz2011Entanglement}, the $n$-th R\'enyi entropy being related to the $2n$-point correlator of the corresponding field. In this work, this result is derived in the more general framework of the Calabrese and Cardy article of 2004. The result is found to reproduce correctly the entropy of primary states of several models: the $XX$, $XXZ$ and excluded-$XXZ$ models in the $c=1$ universality class, and the Ising model in the $c=1/2$ universality class. Predictions for the low-$\ell$ behaviour of the von Neumann entropy are also numerically verified for bosons. We present numerical evidence of the universality of the parity effect of the R\'enyi entropy also for excited states, although further investigation is needed. \\
\indent
Finally, we have studied the R\'enyi entropies of some descendant states in the $XX$ and the Ising model, finding a partial sucess of the analytic approach which indicates that further investigation is required to fully understand the entropies of all the low energy states in critical lattice systems. 

\section{Acknowledgements}

This work was  supported by the Spanish projects FIS2009-11654 and QuiteMad. We acknowledge the IFT (UAM/CSIC) for letting us use the high-performance computing cluster \textit{Hydra}.

\appendix

\section{Entanglement in the $XY$ model \textit{via} correlation matrices} \label{XYfermionization}

We consider the $XY$ model for $N$ spin-1/2 particles with periodic boundary conditions (PBC) in an external $z$ field.

\beq
H_{XY}=\frac{-1}{2}\sum_{j=1}^N\frac{1}{2}\left[(1+\gamma)\sigma^x_j\sigma^x_{j+1}+(1-\gamma)\sigma^y_j\sigma^y_{j+1}\right]+J\sigma_j^z
\label{Hxy}
\eeq
where $\sigma^\alpha$ are the Pauli matrices:

\beq
\sigma^x=\matriz{0}{1}{1}{0},\qquad \sigma^y=\matriz{0}{-i}{i}{0},\qquad \sigma^z=\matriz{1}{0}{0}{-1}
\nonumber
\eeq

This generalizes the Ising (ITF) ($\gamma=1$) and $XX$ ($\gamma=0$, $J=0$) models. Defining the raising and lowering spin operators $\sigma^{\pm}=\frac{1}{2}(\sigma^x\pm i\sigma^y)$ we have:

\beq
H_{XY}=\frac{-1}{2}\sum_{j=1}^N\,\left[\sigma^+_j\sigma^-_{j+1}+\gamma\,\sigma^+_j\sigma^+_{j+1}+\mbox{h.c.}\right]+J\sigma_j^z \nonumber
\eeq
A Jordan-Wigner transformation is now performed to express (\ref{Hxy}) in terms of true fermions $\{c_m^\dag,c_{n}\}=\delta_{m,n}$. They are defined as:

\beq
c_m=\left(\prod_{j<m}\sigma^z_j\right)\,\sigma_m^-
\eeq
and they satisfy 
\beq
c^\dag_Nc_1=-(-1)^{n_\downarrow}\sigma^+_N\sigma_1^-
\label{fermionbc}
\eeq
where $n_\downarrow\equiv N-n_F$ is the number of down spins and 

\beq
n_F=\sum_{j=1}^N\frac{s_j^z+1}{2}
\eeq 
is the number of up spins ($s_j^z$ being the eignevalue of $\sigma_j^z$) and also the number of fermions. Equation (\ref{fermionbc}) determines the boundary conditions in the fermionic formulation of (\ref{Hxy}) and the Hamiltonian reads ($c_{N+1}=(-1)^{n_\downarrow}c_1$):

\barray
H_{XY}=\frac{1}{2}\sum_{j=1}^N\,\left[c^\dag_jc_{j+1}+\gamma\, c^\dag_jc^\dag_{j+1}+\mbox{h.c.}\right]-2Jc^\dag_j c_j \nonumber 
\earray
A Fourier transform is now performed. The $N$ fermionic modes, $\{d_j^\dag,d_{j'}\}=\delta_{j,j'}$ are defined:

\beq
d_j=\frac{1}{\sqrt N}\sum_{m=1}^N e^{-2\myi  k_j m}\,c_m
,
\eeq
(where $k_j \equiv 2\pi j /N$ is the momentum associated with index $j$), in such a way that:

\beq
H_{XY}=\sum_{j\in\Omega}\, A_j\, d_j^\dag\,d_j+i B_j\, \left[d_j^\dag d_{-j}^\dag+ d_j d_{-j}\right]
\label{XYmomentum}
.
\eeq
where we defined $A_j$, $B_j$ :

\beq
A_j=\cos{k_j} -J,\qquad B_j=\frac{\gamma}{2} \sin{k_j}
,
\eeq
and where the set of momentum indexes $\Omega$ is such that the resulting allowed momenta $k_j$, $j\in\Omega$, are those specified later in equation \ref{allowedmomenta}. The Hamiltonian (\ref{XYmomentum}) describes interacting fermions and can be diagonalized throught a Bogolubov transformation. Defining the fermionic modes ($u_j^2+v_j^2=1$):

\barray
b_j^\dag=u_j\, d_j^\dag+i v_j\, d_{-j} \nonumber \\
b_{-j}=u_j\, d_{-j}+i v_j\, d_j^\dag 
\label{Bogolubovt}
,
\earray
and imposing that the Hamiltonian for the $j$, $-j$ modes has the diagonal form $(E_j\,b_j^\dag b_j-E_j\,b_jb_j^\dag)/2$, i.e., that:

\barray
H_{XY}=\frac{1}{2}\sum_j\, E_j\, b_j^\dag b_j-E_j\, b_j b_j^\dag=\sum_j\, E_j\, b_j^\dag b_j-\frac{1}{2} E_j
\label{imposedH}
\earray
we obtain the Bogolubov equations to be satisfied by $u_j$, $v_j$:

\barray
u_j^2-v_j^2=A_j/E_j \nonumber  \\
-u_j v_j = B_j/E_j
\label{Bogoluboveqns}
\earray
and (\ref{imposedH}) with

\beq
E_j^2=\left(\cos{k_j} -J\right)^2+\gamma^2 \sin^2{k_j}
\label{Bogolubovspectrum}
.
\eeq


There is an ambiguity in the sign of $E_j$ in (\ref{Bogolubovspectrum}). Each $j,-j$ being diagonalized independently, one can choose the sign of $E_j$ independently, $E_j=E_{-j}=|E_j|s_j$. With the election $s_j=-1$, the ground state of $H_{XX}$ corresponds to the ground state of the free fermion system (\ref{imposedH}) at half filling, as said in section \ref{XXoverview} (even $N$ is supposed). \\
\indent
In conclusion, via the Bogolubov transformation (\ref{Bogolubovt},\ref{imposedH},\ref{Bogoluboveqns}), we have expressed the $XY$ problem in terms of a free fermion problem, in such a way that the eigenstates of $H_{XY}$ in this free fermionic formulation are:

\beq
\prod_{q\in \mn K}b_{q}^\dag|0\>
\eeq
and that the spectra of $H_{XY}$ is $\sum_{j\in \mn K}E_j-K$, being $|0\>$ the fermion vaccum and $\mn K\subset \Omega$ is one of the a set of $n_F$ integers or half-integers characterizing the state, such that $k_j=2 \pi\,j/N$ is the momentum corresponding to index $j\in\mn K$. The set of allowed momenta depends on the boundary conditions of the fermionic problem, which are in their turn determined by the parity of $n_\downarrow$ (\ref{fermionbc}):

\begin{IEEEeqnarray}{l}
\stackrel{N-n_F \mbox{ even}}{\mbox{(APBCs)}} \left\{
\begin{array}{rl} 
N \mbox{ even:} & N\phi=\pm\pi,\pm 3\pi,\ldots,\pm (N-1)\pi\\
N \mbox{ odd:} & N\phi=\pm \pi,\pm 3\pi,\ldots,\pm  (N-2)\pi,N\pi
\end{array}  \right. \nonumber \\
\stackrel{N-n_F \mbox{ odd}}{\mbox{(PBCs)}} \left\{
\begin{array}{rl} 
N \mbox{ even:} & N\phi=0,\pm 2\pi,\pm 4\pi,\ldots,\pm (N-2)\pi,N\pi\\
N \mbox{ odd:} & N\phi=0,\pm  2\pi,\pm 4\pi,\ldots,\pm (N-1)\pi
\end{array}  \right. 
\label{allowedmomenta}
.
\end{IEEEeqnarray}
 
\subsection{Correlation matrix in terms of Majorana fermions}

The $2N$ spatial Majorana modes $\bar c_r$ are defined:

\barray
c_m &  = & \frac{\bar c_{2m-1}+i\bar c_{2m}}{2} \nonumber \\
\bar c_{2m-1}& = & c_m^\dag+c_m \nonumber \\
\bar c_{2m}  & =& i(c_m^\dag-c_m) 
\label{Majorana}
\earray
they satisfy Majorana anticommutation rules: $\{\bar c_r,\bar c_s\}=2\delta_{r,s}$. We want to construct the correlation matrix $\<\bar c_r \bar c_s\>_\mn K$ for a given state defined by $\mn K$. We first express the $\bar c$'s in terms of the $d$'s, then the  $\bar c$'s in terms of $b$'s using the inverse Bogolubov transformation (\ref{Bogolubovt}). Finally, the set $\mn K$ determines the occupancies of the free fermions: $\<b_j^\dag b_{j'}\>_\mn K=\delta_{j,j'}$ if $j,j'\in\mn K$, and zero otherwise, and $\<b_j^\dag b^\dag_{j'}\>=\<b_j b_{j'}\>=0$, and one gets for $\<\bar c_r \bar c_s\>$:



\beq
\<\bar c_r \bar c_s\>=\delta_{rs}+\myi\Gamma_{rs}
\label{correlatorMajorana}
\eeq
with $\Gamma \in \mathcal{M}_{2N}(\mathbb C)$:

\barray
\Gamma=\left(\begin{array}{cccc}
\Pi_0   & \Pi_1   &    \ldots & \Pi_{N-1}\\
\Pi_{-1}   & \Pi_0   &    \ldots & \vdots\\
\vdots  &         &    \ddots &   \\
\Pi_{-N+1} & \ldots &    & \Pi_0
 \end{array}\right)
,
\earray

\barray
\Pi_m=\left(\begin{array}{cc}
g^{(1)}_m   & g^{(2)}_m \\
g^{(3)}_m & g^{(1)}_m \\
 \end{array}\right)
\earray
and where:

\barray
g^{(1)}_m=\frac{1}{N}\left[\sum_{k\in \mn K} e^{i\phi_k m} + \sum_{k\notin \mn K} e^{-i\phi_k m}\right] \nonumber \\
g^{(2)}_m=\frac{i}{N}\left[\sum_{k\in \mn K} -e^{-i\phi_k m} (u_k-iv_k)^2 + \sum_{k\notin \mn K} e^{i\phi_k m} (u_k+iv_k)^2\right] \nonumber \\
g^{(3)}_m=\frac{i}{N}\left[\sum_{k\in \mn K} e^{-i\phi_k m} (u_k+iv_k)^2 - \sum_{k\notin \mn K} e^{i\phi_k m} (u_k-iv_k)^2\right]
.
\earray

From the antisymmetry of $\<\bar c_r \bar c_s\>$ one can see that $g^{(2)}_m=-g^{(3)}_{-m}$ and $g^{(1)}_m=-g^{(1)}_{-m}$. For the ground state, it is $\mn K=-\mn K$ and the same for the complementary of $\mn K$ (this is not true for a general state). One now uses this condition: $\sum_j \to \sum_{-j}$ in the expression for $g_m$, and also the fact that, for the ground state, $E_j=-|E_j|$ if $j\in \mn K$, $E_k=|E_k|$ otherwise, and we obtain:

\barray
g_m=\frac{1}{N}\sum_{j \mbox{ all}} \frac{A_j-2\myi B_j}{|E_j|} \,e^{-\myi \phi_j m}
\label{myg}
\earray
which in the continuum limit becomes equation (3.47) in ~\cite{Latorre2004Ground}, of which (\ref{correlatorMajorana}) is a generalization for arbitrary states characterized by $\mn K$ and for finite-size systems.\\





\subsection{Computation of entanglement}
\label{entanglementviaMajorana}

We are interested in the entanglement between a spatial partition containing the first $\ell$ sites of the system, and the rest of the system. Being $\bar c_{2m}$, $\bar c_{2m -1}$ related to the $m$-th site, the reduced density matrix of the subsystem defined by $1\le \ell \le N$ is encoded ~\cite{Latorre2004Ground} in a $2\ell$-block of matrix $\Gamma$, $\Gamma_\ell\in \mathcal{M}_{2\ell}(\mathbb C)$:

\barray
\Gamma_\ell= \left(\begin{array}{cccc}
\Pi_0   & \Pi_1   &    \ldots & \Pi_{\ell-1}\\
\Pi_{-1}   & \Pi_0   &    \ldots & \vdots\\
\vdots  &         &    \ddots &   \\
\Pi_{-\ell+1} & \ldots &    & \Pi_0
 \end{array}\right)
\label{GammaL}
.
\earray
In particular, the entanglement is computed in the following way. As $(\<\bar c_r \bar c_s\>)_{rs}$, $\Gamma_\ell$ is an antisymmetric matrix whose eigenvalues are complex coniugated pure imaginary numbers $\pm \myi \nu_r$, $\nu_r$ being real and depending on $\ell$. Such a diagonal matrix can be transformed into a block-diagonal form:

\beq
\tilde \Gamma_\ell = \bigoplus_{r=1}^\ell \nu_r \left(\begin{array}{cc}0&1\\-1&0\end{array} \right)
.
\eeq
Now suppose that $\bar a_m$ is the basis in which this happens, i. e., $\<\bar a_r \bar a_s\>=\delta_{rs}+\myi(\tilde \Gamma_\ell)_{rs}$. It is easy to see that the corresponding true fermions $a_m=\bar a_{2m-1}+\myi\bar a_{2m}$ are in a product of uncorrelated states: 

\beq
\<a_m a_n\>=0, \qquad \<a^\dag_m a_n\>=\delta_{mn}\frac{1+\nu_m}{2}
.
\eeq
This product being uncorrelated, the reduced correlation matrix of the $\ell$-block is a product $\otimes_m^\ell \varrho_m$ of correlation matrices of single-site blocks $\varrho_j=p_j\,a_j^\dag|0\>\<0|a_j+(1-p_j)|0\>\<0|$, $p_j=(1+\nu_j)/2$. The entanglement entropy $S(\ell)$ of the $\ell$-lengthed block is, hence, the sum of the entropies of the $\varrho_m$'s:

\beq
S(\ell)=\sum_{j=1}^\ell H_2(\frac{1+\nu_j}{2})
\label{Sviareduced}
.
\eeq
being $H_2(x)=-x\ln x-(1-x)\ln (1-x)$. The strategy is hence to construct the matrix (\ref{GammaL}) numerically for a given state caracterized by the set $\mn K$. By diagonalyzing $\Gamma_\ell$ we obtain $\nu_r$ and finally the entropy via (\ref{Sviareduced}). In the same way, the R\'enyi entropy can be computed:

\beq
S_n(\ell)=\sum_{j=1}^\ell \ln \left[ {p_j}^n+(1-p_j)^n\right]
.
\eeq
The total time cost of the operation being $O(\ell^3)$.

\bibliographystyle{apsrev}
\bibliography{berganza-paper1}

\end{document}